\documentclass[%
reprint,
superscriptaddress,
 twocolumn,
 amsmath,amssymb,
 aps,
 prb,
 longbibliography,
floatfix,
]{revtex4-2}

\usepackage{graphicx}
\usepackage{color}
\usepackage{dcolumn}
\usepackage{bm}
\usepackage[normalem]{ulem} 
\usepackage{enumerate}
\usepackage{physics}
\usepackage[colorlinks=true,citecolor=blue,linkcolor=blue,urlcolor=blue]{hyperref}
\usepackage{cleveref}
\usepackage{float}
\usepackage{multirow}
\usepackage{makecell}
\usepackage{soul}
\usepackage{cancel}
\usepackage{tabularx}
\newcolumntype{Y}{>{\centering\arraybackslash}X}

\newcommand{\Eo}[0]{E_\textrm{orb}} 
\newcommand{\kB}[0]{k_\textrm{B}} 
\newcommand{\omegaq}[0]{\omega_\textrm{q}} 
\newcommand{\fL}[0]{f_\textrm{L}} 
\newcommand{\muB}[0]{\mu_\textrm{B}} 
\newcommand{\ld}[0]{\ell_\textrm{d}} 

\begin{document}

\title{Acoustic phonons, spin-phonon coupling and spin relaxation via the lattice reorientation mechanism in hexagonal germanium nanowires}
\author{Baksa Kolok}
 \email{kolokba@edu.bme.hu}
\affiliation{%
 Department of Theoretical Physics, Institute of Physics, Budapest University of Technology and Economics, Műegyetem rkp. 3., H-1111 Budapest, Hungary
}%
\affiliation{HUN-REN-BME-BCE Quantum Technology Research Group, Budapest University of Technology and Economics, Műegyetem rkp. 3., H-1111 Budapest, Hungary}
\author{György Frank}
\affiliation{%
 Department of Theoretical Physics, Institute of Physics, Budapest University of Technology and Economics, Műegyetem rkp. 3., H-1111 Budapest, Hungary
}%
\author{András Pályi}%
\affiliation{%
 Department of Theoretical Physics, Institute of Physics, Budapest University of Technology and Economics, Műegyetem rkp. 3., H-1111 Budapest, Hungary
}%
\affiliation{HUN-REN-BME-BCE Quantum Technology Research Group, Budapest University of Technology and Economics, Műegyetem rkp. 3., H-1111 Budapest, Hungary}

\date{\today}

\begin{abstract}
Spin relaxation via electron-phonon interaction is an important decoherence mechanism for spin qubits. 
In this work, we study spin relaxation in hexagonal (2H) germanium, a novel direct-gap semiconductor showing great potential to combine highly coherent spin qubits with optical functionality. 
Focusing on electrostatically defined quantum dots in hexagonal germanium nanowires, we (i) identify geometries where spin qubit experiments are feasible, (ii) compute the nanowire phonon modes, and (iii) describe spin relaxation of hole spin qubits due to phonon-induced lattice reorientation, a direct spin-phonon coupling mechanism that is absent in cubic semiconductors typically used for spin qubits (GaAs, cubic Si, cubic Ge). 
We obtain the spin relaxation time as a function of nanowire cross section, quantum dot confinement length, and magnetic field.
For realistic parameters, we find relaxation times above 10 ms, and reveal that the magnetic field direction maximizing the relaxation time depends on the qubit Larmor frequency.
Our results facilitate the design of nanowire quantum dot experiments with long qubit relaxation times.
\end{abstract}

\maketitle

\section{Introduction}
\label{sec:intro}

 Hexagonal germanium has a direct bandgap \cite{Fadaly2020Direct-bandgapAlloys, Rodl2019AccurateApplications}, paving the way for optoelectronic applications of germanium and germanium-silicon alloys, potentially leading to significant advances in both conventional and quantum information technology \cite{Borlido2023EnsembleAlloys, vanTilburg2024StimulatedNanowires, JacquesI.Pankove1975OpticalSemiconductors, Soref2015EnablingCommunications, Thomson2016RoadmapPhotonics, Poulton2017CoherentArrays, Wang2017IIIV-on-SiliconRange, Munoz2017SiliconApplications, Wang2018MultidimensionalOptics, Shen2019SiliconSystems}. 
 Although cubic silicon and germanium serve as host materials for standard semiconductor devices, the potential in cubic silicon and germanium for these optoelectronic applications is limited due to their indirect bandgap \cite{Pavesi2000OpticalNanocrystals, Han2001OpticalWaveguide, Rong2005AnLaser}. 
 Moreover, both germanium and silicon hold significant potential for spin-based quantum computing---whether in planar  \cite{
 Burkard2023SemiconductorQubits,
 Scappucci2020TheRoute,
 Chatterjee2021SemiconductorPractice,
 Petta2005CoherentDots, 
 Koppens2006DrivenDot, 
 Nowack2007CoherentFields,
 Pla2012ASilicon,
 Kim2014QuantumQubit,
 Veldhorst2017SiliconComputer, 
 Vandersypen2017InterfacingCoherent, 
 Crippa2018ElectricalQubits,
 Hendrickx2021AProcessor,  
 Jirovec2021AGe, 
 Gonzalez-Zalba2021ScalingTechnology,
 Takeda2022QuantumQubits, 
 Philips2022UniversalSilicon,
 Xue2022QuantumThreshold,
 Wang2024PursuingDot} or nanowire structures
 \cite{Brauns2016AnisotropicDots,
 Vukusic2017FastReflectometry,
 Froning2021StrongDots,
 Froning2021UltrafastFunctionality,
 Camenzind2022AKelvin}---thanks to their low abundance of nuclear spins and compatibility with standard CMOS technology. 
 Therefore, hexagonal germanium and silicon-germanium materials are particularly significant for developing optical and electrical quantum dots (QD) on the same chip, acting as an interface between stationary (spin-based) and mobile (photon-based) qubits, in addition to other optoelectronic devices.

Phonons are fundamental for the thermal, electrical, and optical behavior of semiconductor materials, making them a key focus in the design and analysis of semiconductor devices \cite{Yu2010FundamentalsSemiconductors,Solyom2007FundamentalsSolids}. In particular, in the case of quantum processors, phonons can significantly impact the performance of qubits by interacting with spin qubits, leading to decoherence and loss of quantum information \cite{Khaetskii2000SpinDots, Elzerman2004Single-shotDot, Amasha2008ElectricalDot, Hu2012HoleQubits, Maier2013TunableDots, Tahan2014RelaxationDots, Boross2016ControlStep}. 

\begin{figure}[t]
    \centering
    \includegraphics[width=.9\linewidth]{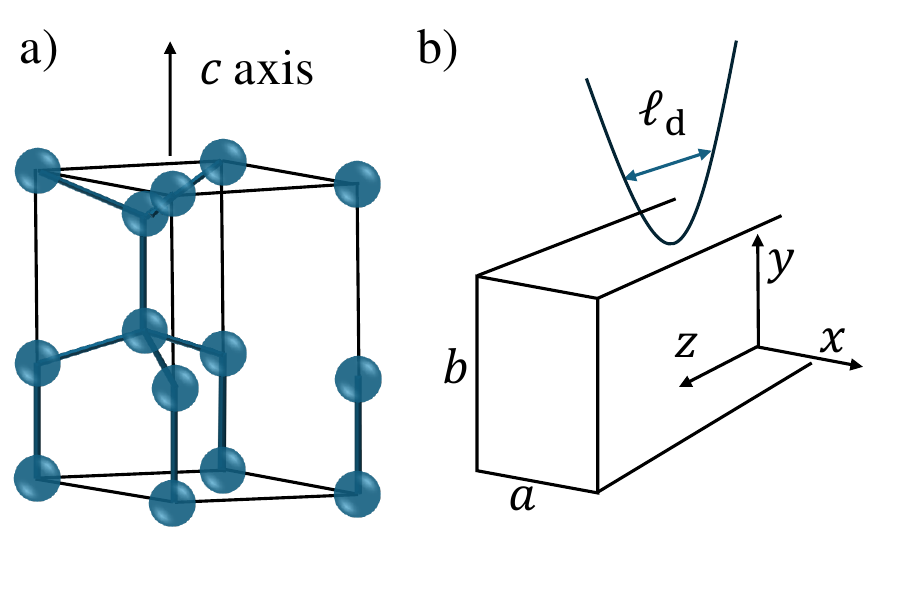}
    \caption{{The geometry of the rectangular cross-section hexagonal germanium nanowire compared to the crystal orientation.} (a) The unit cell of hexagonal germanium. The $c$ axis is the sixfold screw axis of the space group of the crystal. (b) The nanowire geometry. The cross-section is rectangular and the nanowire is infinite in the $z$ direction, in which direction the confinement is assumed to be parabolic with $\ld$ confinement length. The growth direction of the nanowire is perpendicular to the $c$ axis of the crystal.} 
    \label{fig:nw_geom}
\end{figure}

This work focuses on the device geometry shown in Fig.~\ref{fig:nw_geom}, i.e.~a single-crystal nanowire of hexagonal germanium grown perpendicular to the axis of hexagonal symmetry, that is, the $c$ axis of the crystal, with rectangular cross section, which is the geometry of the state-of-the-art hexagonal germanium nanowires \cite{Li2023HexagonalComposition, Lamon2025DimensionNanowires}. In addition, we model a harmonic confinement realized in the longitudinal direction by electrical gates. 

We explore the conditions required for the implementation of efficient electrical QDs within these nanowires in Sec.~\ref{sec:efficientQD}. These estimates establish the framework for designing and implementing quantum devices with hexagonal germanium nanowires.

We analyze the behavior of acoustic phonons in Sec.~\ref{sec:phonons}, which is essential to further analyze the potential applications of such hexagonal germanium nanowires as host material for conventional semiconductor devices as well as for quantum applications. 

In Sec.~\ref{sec:relaxation}, we model the phonon-induced lattice reorientation mechanism, a direct spin-phonon coupling absent in cubic materials, originating from the anisotropy of the hexagonal crystal structure. We calculate the resulting relaxation rates of semiconductor spin qubits hosted in hexagonal germanium nanowires caused by phonons via the lattice reorientation mechanism.
We obtain the spin relaxation time as a function of nanowire cross section, quantum dot confinement length, and magnetic field. 
For realistic parameters, we find relaxation times above 10 ms, and reveal that the magnetic field direction maximizing the relaxation time depends on the qubit Larmor frequency.
Our findings indicate that hexagonal germanium nanowires carry the potential of future applications in quantum information processing as highly coherent spin-photon interfaces, and help the design of spin-qubit experiments using nanowire quantum dots.

\section{Spin-qubit functionality imposes constraints on the geometry of the quantum dot}
\label{sec:efficientQD}

Our main goal in this work is to describe the decoherence of quantum-dot spin qubits in hexagonal germanium nanowires.
We will focus on a particular type of nanowire that has already been synthesized \cite{Li2023HexagonalComposition}. 
However, quantum-dot spin qubits in such wires have not yet been realized, hence it is an important preliminary task to identify the geometric constraints on such a nanostructure which need to be fulfilled to reach that milestone.
This is what we do in the rest of this section. 

The type of nanowires we consider is grown in the direction perpendicular to the $c$ axis, and has a rectangular cross section, with one of the sides of the rectangle facing  the $c$ axis.
This geometry is illustrated in Fig.~\ref{fig:nw_geom}.
The side lengths of the rectangular cross section are denoted by $a$ and $b$, respectively.
We assume that these may vary from a few nanometers to a few hundreds of nanometers, and will use $a = b = 30\, \mathrm{nm}$ as a representative example. 

A central quantity in a spin-qubit experiment is the orbital level spacing $E_\mathrm{orb}$: this should be greater than the energy scale $\hbar \omega_q$ of the qubit Larmor frequency, and it should also well exceed the thermal energy scale $k_\mathrm{B} T$.
(Here, $\hbar$ is the reduced Planck constant and $k_\mathrm{B}$ is the Boltzmann constant.)
We focus on the case where a single carrier is confined along the nanowire axis $z$ with a parabolic electrostatic potential induced by gate electrodes, as illustrated in Fig.~\ref{fig:nw_geom}(b).
We model this system with hard-wall confinement in the $x$ and $y$ directions, and harmonic confinement in the $z$ direction, with a ground-state broadening of $\ld$. 
Also, we assume that the physical length of the wire is much greater than $\ld,\, a$ and $b$.
In this case, the orbital splitting from the ground state reads as
\begin{equation}
    \Eo = \min\left\{ \frac{3h^2}{8m_x a^2},\frac{3h^2}{8m_y b^2}, \frac{\hbar^2}{m_z\ld^2}\right\},
\end{equation}
where $m_x,\,m_y$ and $m_z$ are the effective masses corresponding to the given direction. 
We have collected the effective mass parameters for the lowest conduction band and the highest valence band in Table \ref{tab:bandparamaters}.

\begin{table}[t]
\footnotesize
\centering
\caption{Effective mass and $g$ factor values of the highest valence band and the lowest conduction band of bulk 2H germanium.\cite{Pulcu2024MultibandGermanium}.}
\label{tab:bandparamaters}
{\setlength{\tabcolsep}{1pt}
\begin{tabularx}{\linewidth}{lYcc}
\hline\hline 
\noalign{\vskip 3pt}
band                &  {\spaceskip=2pt direction with respect to~$c$ axis} & $m_\text{eff}/m_\text{e}$ & $g$ factor \vspace{1mm} \\ \hline 
\noalign{\vskip 3pt}
valence band & perpendicular ($\perp$)   & 0.08                    & 2          \\
& parallel ($\parallel$)    & 0.51                     & -18.225    \\ 
conduction band & perpendicular ($\perp$)   & 0.12                     & 2          \\ 
& parallel ($\parallel$)    & 1                     & 2          \\[2pt] \hline\hline
\end{tabularx}}
\end{table}

\begin{figure}[!b]
    \centering
    \includegraphics[width=.9\linewidth]{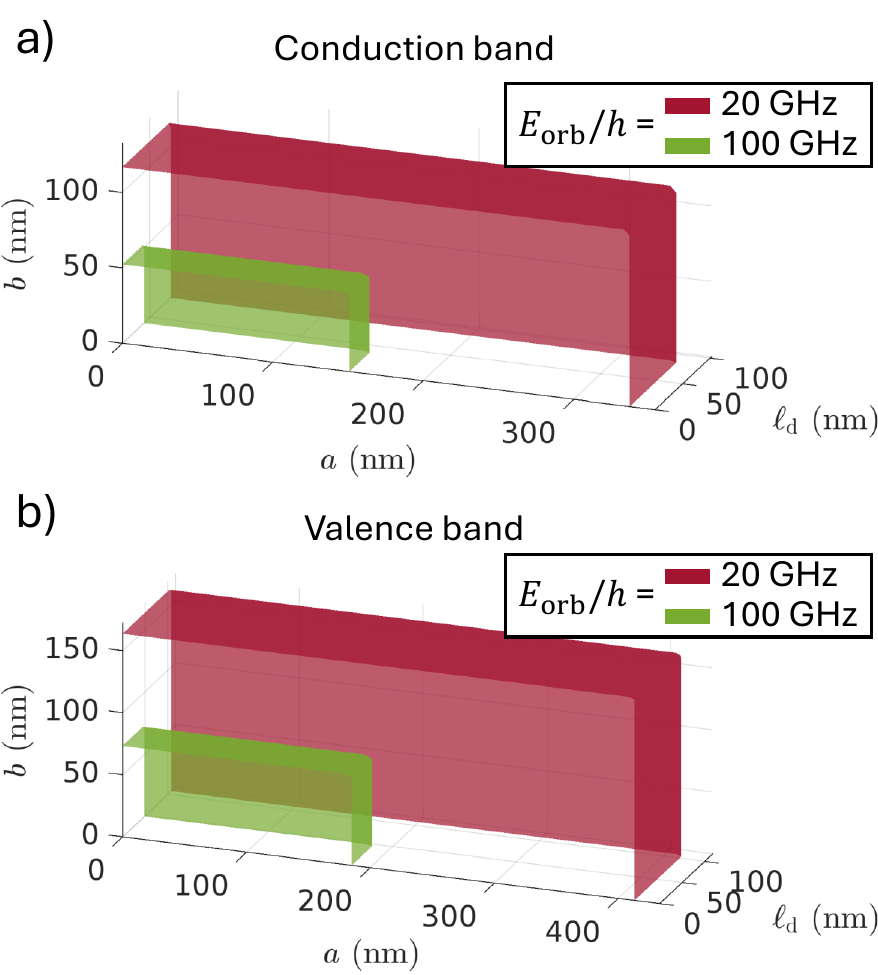}
    \caption{{Orbital level spacing frequencies for a quantum dot hosted in a rectangular cross-section hexagonal germanium nanowire.} The $x$ and $z$ axes show the width of the nanowire, the y axis shows the width of the harmonic confinement in the longitudinal direction. Green surface corresponds to $E_\mathrm{orb}/h =  100\, \mathrm{GHz}$ orbital splitting, the red to the 20 GHz ($\approx1$ K thermal broadening) orbital splitting. The first excited orbital state needs to be well separated from the ground state compared to the thermal broadening ($\kB T$)  and the qubit energy splitting ($\hbar\omegaq$). Inside the green surface ($\Eo/h > 100$ GHz), we expect the usual spin-qubit functionality of the system.}
    \label{fig:orb_splitting}
\end{figure}

Figure \ref{fig:orb_splitting} illustrates how the orbital level spacing $E_\mathrm{orb}$ depends on the three confinement lengths $a$, $b$, and $\ld$, for the conduction band (a) and the valence band (b).
In both panels, the green (red) surface correspond to confinement length values that set the orbital level spacing to 
$E_\mathrm{orb}/h = 100\, \mathrm{GHz}$ ($E_\mathrm{orb}/h = 20\, \mathrm{GHz}$).
Since the Larmor frequency $\omega_\mathrm{q}$ of a spin qubit is typically at most in the range of tens of GHz \cite{Burkard2023SemiconductorQubits}, and recent experiments suggest to decrease it even below 1 GHZ \cite{Hendrickx2024Sweet-spotSensitivity}, one can avoid the overlap of the Zeeman sublevels of the ground-state orbital and the first excited orbital by keeping the orbital level spacing above 100 GHz. 
This is fulfilled for strong enough confinement, shown as the rectangular cuboid volume within the green boundary in Fig.~\ref{fig:orb_splitting}.
We tabulate the confinement length values corresponding to the boundaries of this green volume in Table \ref{tab:parameterupperbounds}.

Note also that an orbital level spacing of 100 GHz corresponds to a temperature of 5 K, hence this orbital level spacing ensures efficient thermalization to the ground-state orbital at the subkelvin temperatures that are typical in spin-qubit experiments. 

In conclusion, the green rectangular cuboid volume in Fig.~\ref{fig:nw_geom} indicates the confinement lengths that should be achieved in fabrication, in order to ensure usual quantum-dot spin qubit functionality.

\section{Acoustic phonons in a nanowire} \label{sec:phonons}

In this section, we calculate the acoustic phonon modes, and their dispersion relations, in a hexagonal germanium nanowire. 
We focus on the geometry defined in the previous section, and use continuum elasticity theory.
Following the standard procedure \cite{Vlasov1961Thin-walledBeams, Gurtin1973, Landau1986TheoryElasticity, Davi1996DynamicsRods}, for simplicity we assume that the nanowire is infinitely long and floating. 
The results we derive here are essential to compute the phonon-related relaxation times for spin qubits.

\begin{table}[]
\centering
\caption{The upper bounds for the cross-section widths ($a,b$) and for the harmonic confinement ground-state spreading ($\ld$), to keep the orbital splitting above 100 GHz. 
Below the upper bounds we expect the usual spin-qubit functionality of the QD.
These numbers correspond to the parameters of the green boxes in Fig.~\ref{fig:orb_splitting}.}
\label{tab:parameterupperbounds}
{
\begin{tabularx}{\linewidth}{l Y Y c}
\hline\hline 
\noalign{\vskip 3pt}
band                & $a$ (nm) & $b$ (nm) & $\ld$ (nm) \\[2pt] \hline
\noalign{\vskip 3pt}
valence band &  184.7  & 73.2       & 48          \\ 
conduction band &  150.8  & 52.3       & 39.2          \\ 
\noalign{\vskip 3pt} \hline\hline
\end{tabularx}}
\end{table}

\subsection{Continuum elasticity theory of the nanowire: equation of motion and boundary conditions}

The general equation of motion in continuum elasticity theory for a homogeneous piece of material reads \cite{Landau1986TheoryElasticity}:
\begin{equation}\label{eq:EoM}
    \rho \Ddot{u}_i(\mathbf r, t) = \sum_{j \in \{x,y,z\}}\frac{\partial \sigma_{ij}(\mathbf r,t)}{\partial r_j}, \qquad \forall i\in\{x,y,z\},
\end{equation}
where $\mathbf{r} = (r_x,r_y,r_z) = (x,y,z)$ is the position in the material,
$\mathbf{u}(\mathbf r,t)$ is the displacement field and $\sigma_{ij}(\mathbf r,t)$ is a matrix element of the stress tensor, and $\rho$ is the density of the material. 
We are interested in the small displacement limit, where relation between the strain and stress tensors is linear (Hooke's law), and is given by the stiffness tensor $C$.
This relation reads \cite{Landau1986TheoryElasticity}:
\begin{equation}
\label{eq:hooke}
    \sigma_{ij} = \sum_{kl} C_{ijkl} \epsilon_{kl},
\end{equation}
where $\epsilon$ is the strain tensor defined as
\begin{equation}
\label{eq:straindef}
    \epsilon_{ij} = \frac{1}{2}\left(\frac{\partial u_i}{\partial r_j} + \frac{\partial u_j}{\partial r_i}\right).
\end{equation}
Using Eqs.~\eqref{eq:hooke} and
\eqref{eq:straindef}, the equation of motion \eqref{eq:EoM}  is expressed as
\begin{equation} \label{eq:christoffel}
    \rho\Ddot{u}_i(\mathbf{r},t)  = \sum_{jkl}C_{ijkl}\partial_j\partial_k u_l,
\end{equation}
where we used that the stiffness tensor is symmetric under the exchange of the last two indices.

We assume that the nanowire is a single-crystal material from hexagonal germanium.
Therefore, stiffness tensor in the Voigt notation \cite{Voigt1966LehrbuchKristallphysik} in a coordinate system, where the $c$ axis is parallel with the $y$ axis (see Fig.~\ref{fig:nw_geom}), reads as \cite{Landau1986TheoryElasticity}
\begin{equation} \label{eq:Chex}
    C =
    \begin{pmatrix}
        C_{11} & C_{13} & C_{12} & 0 & 0 & 0 \\
        C_{13} & C_{33} & C_{13} & 0 & 0 & 0 \\
        C_{12} & C_{13} & C_{11} & 0 & 0 & 0 \\
        0 & 0 & 0 & C_{66} & 0 & 0 \\
        0 & 0 & 0 & 0 & C_{44} & 0 \\
        0 & 0 & 0 & 0 & 0 & C_{44} 
    \end{pmatrix},
\end{equation}
and $C_{66} = \frac{C_{11} - C_{12}}{2}$. The numerical values of the matrix elements of $C$ were calculated for hexagonal germanium using \textit{ab initio} methods. We have summarized such results from three different references in Table \ref{tab:elastic_const}.
For our numerical calculations, we used the values from Ref.~\cite{Mayengbam2023TheoreticalEngineering}.

\begin{table}[t]
\centering
\caption{{Elements of the stiffness tensor from \textit{ab initio} calculations for hexagonal germanium in units of GPa.} In our numerical calculations we use the values from the first column (Ref.~\cite{Mayengbam2023TheoreticalEngineering}).}
\label{tab:elastic_const}
\begin{tabularx}{\linewidth}{lYYc}
\hline\hline
\noalign{\vskip 3pt}
& Ref.~\cite{Mayengbam2023TheoreticalEngineering} & Ref.~\cite{Suckert2021EfficientGermanium} & Ref.~\cite{Wang2003AbGe} \\[2pt] \hline
\noalign{\vskip 3pt}
$C_{11}$ & 143.2        & 124                               & 155.6        \\ 
$C_{12}$ & 40.4         & 53.7                              & 37.5         \\ 
$C_{13}$ & 25.1         & 22.8                              & 27.7         \\ 
$C_{33}$ & 168.7        & 159.4                             & 169.3        \\ 
$C_{44}$ & 41.1         & 39.1         & 41.1         \\ 
$C_{66}$ & 51.4         & 35.2         & 59.1         \\[2pt] \hline \hline
\end{tabularx}
\end{table}

Here, we consider phonons in a nanowire, the latter being translationally invariant along the $z$ direction. 
Hence, we describe wave-like solutions that propagate along the $z$ direction, and have harmonic time dependence.
This consideration implies that the equation of motion will be solved by a displacement field of the following form:
\begin{equation}
\label{eq:trial}
    \mathbf{u}(\mathbf{r},t)
    = 
    \mathbf{w}^q(x,y) e^{i(qz-\omega t)},
\end{equation}
where $q$ is a wave number and $\omega$ is an angular frequency.

Substituting the trial solution of Eq.~\eqref{eq:trial} into Eq.~\eqref{eq:christoffel}, we obtain the following partial differential equation: 
\begin{equation} \label{eq:reducedChristoffel}
    -\rho \omega^2 \mathbf{w}^q (x,y) = \mathbf{D}_\textrm{A} (q) \mathbf{w}^q(x,y),
\end{equation}
where $\mathbf{D}_\textrm{A} (q)$ is a differential operator which reads:
\begin{widetext}
\begin{align} \label{eq:DA}
    \mathbf{D}_\textrm{A} (q) &= \begin{pmatrix}
        C_{11}\partial_x^2 + C_{44}\partial_y^2 - q^2C_{66} & (C_{13} + C_{44})\partial_x\partial_y & iq(C_{12} + C_{66})\partial_x\\ 
        (C_{13} + C_{44})\partial_x\partial_y & C_{44}\partial_x^2 + C_{33}\partial_y^2 - q^2C_{44} & iq(C_{13} + C_{44})\partial_y\\ 
        iq(C_{12} + C_{66})\partial_x & iq(C_{13} + C_{44})\partial_y & C_{66}\partial_x^2 + C_{44}\partial_y^2 - q^2C_{11}
    \end{pmatrix}.
\end{align}
\end{widetext}

Besides the differential equation \eqref{eq:reducedChristoffel}, we also need to specify the boundary conditions for the nanowire. 
We consider a floating nanowire. 
This implies that the stress components perpendicular to the surface of the wire need to vanish. 
We consider a rectangular cross-section, therefore we have four boundary facets. 
The facets are aligned with the $x$ and $y$ axis, see Fig.~\ref{fig:nw_geom}, therefore the boundary conditions for the $x$ and $y$ facets respectively read as
\begin{subequations}\label{eqs:boundary}
\begin{align}
    \sigma_{xx} = \sigma_{xy} = \sigma_{xz} = 0, \quad \textrm{at} \quad x = \pm \frac{a}{2},\\
    \sigma_{yx} = \sigma_{yy} = \sigma_{yz} = 0, \quad \textrm{at} \quad y = \pm \frac{b}{2}.
\end{align}
\end{subequations}
Substituting the form of the displacement field from Eq.~\eqref{eq:trial}, we get three equation for each facet, for the facets aligned with the $x$ axis they read as 
\begin{subequations}
\begin{align}
    C_{11} \partial_{x} w^q_{x} (x,y) +  C_{13} \partial_{y} w^q_{y} (x,y) - q^2 C_{12} w^q_{z} (x,y) &= 0,\\
    C_{13} [\partial_{y}w^q_{x} (x,y) +  \partial_x w^q_{y} (x,y)] &= 0,\\
    C_{12} [iq w^q_{x} (x,y) +  \partial_x w^q_{z} (x,y)] &= 0,
\end{align}
\end{subequations}
and similar equations can be written up for the facets aligned with the $y$ axis.

\subsection{Analytical approximate solutions in the long-wavelength limit} \label{subsec:analdisp}

\begin{figure}[]
    \centering
    \includegraphics[width=\linewidth]{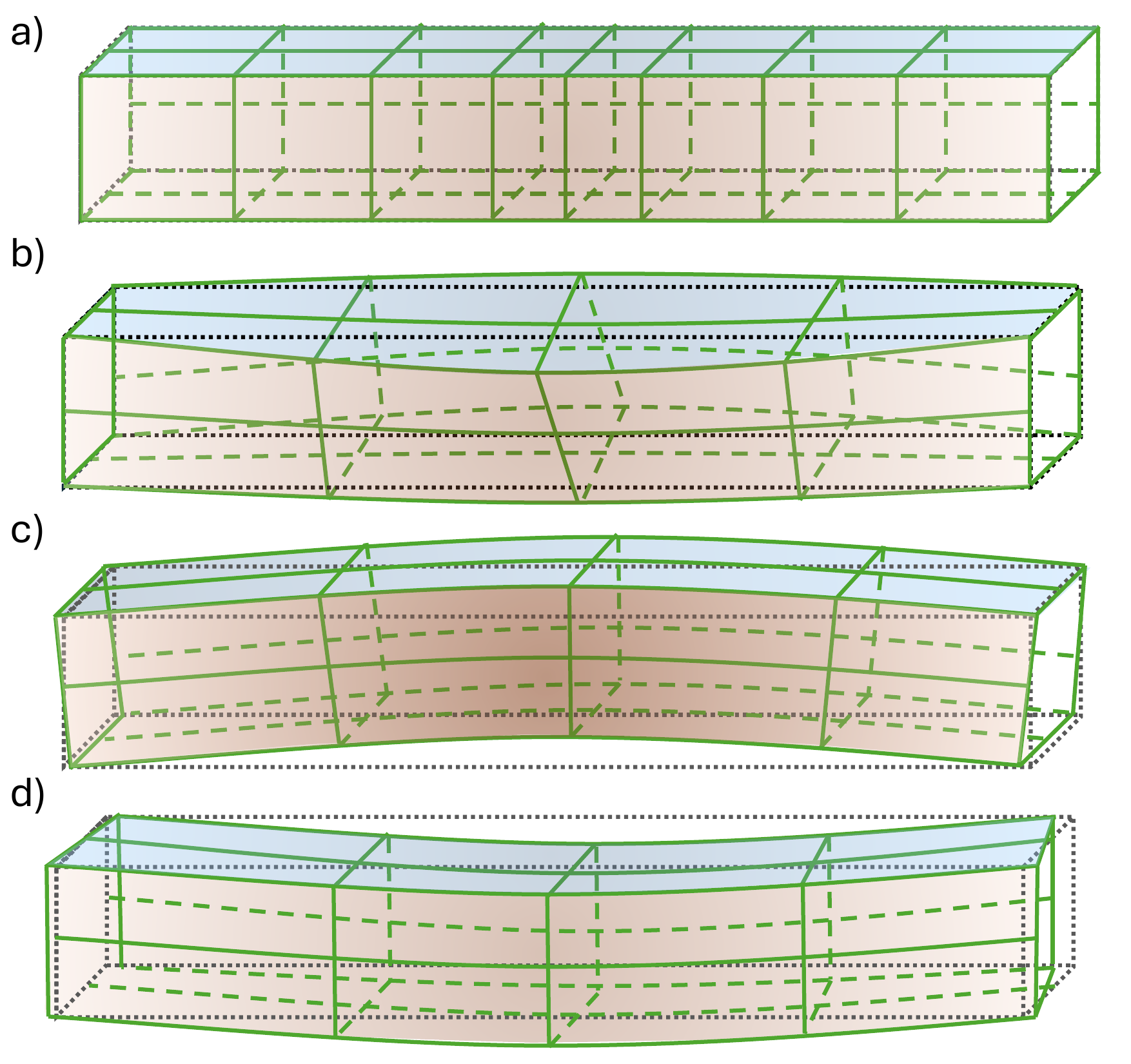}
    \caption{{Shape of the phonon modes in a hexagonal germanium nanowire.} Four gapless phonon modes of the hexagonal nanowire: (a) longitudinal, (b) torsional, (c) $x$ directional flexural and (d) $y$ directional flexural.}
    \label{fig:modes}
\end{figure}

First, we consider the long-wavelength limit, i.e., the case when the wave length $2\pi/q$ of a mode is much greater then the cross-section diamater $d$ of the wire, which we define as $d = \max{(a,b)}$. 
In particular, we focus on the modes that have $\omega = 0$ in the $q = 0$ limit. 

At $q=0$, the displacement field is translationally invariant along the wire, and the zero-frequency modes must not induce any elastic deformation.
This condition restricts the allowed displacements to six infinitesimal rigid-body motions of the cross-section: three translations and three rotations. Among these, the three translations and the rotation around the wire axis (torsion) leave the strain tensor identically zero and thus correspond to zero-energy modes \cite{Vlasov1961Thin-walledBeams, Gurtin1973, Landau1986TheoryElasticity, Davi1996DynamicsRods}. In contrast, the other two rotations (around axes perpendicular to the wire) induce shear deformation and are therefore gapped.

The detailed description and derivation of the equations of motion in the long-wavelength approximation, relying on previous results \cite{Vlasov1961Thin-walledBeams, Gurtin1973, Landau1986TheoryElasticity, Davi1996DynamicsRods}, are found in App.~\ref{app:thin_rod}. 
Here, we state only the solutions for the dispersion relations and the displacement.

The solution of Eq.~\eqref{eq:reducedChristoffel}
for the longitudinal mode, that is, the  shape of the mode and the dispersion relation, in first order of $qd$, reads as
\begin{subequations}
    \begin{align} \label{eq:polL}
        \mathbf{w}^{q}_\textrm{L}(x,y) = A_\textrm{L}[\hat{\mathbf{z}} - iq(x\nu_x \hat{\mathbf{x}} + y\nu_y\hat{\mathbf{y}})],\\
        \omega_\textrm{L}(q) =  \sqrt{\frac{E}{\rho}} q,
    \end{align}
\end{subequations}
where $A_\textrm{L}$ is a scalar that describes the amplitude of the mode, and $E$ is Young's modulus, $\nu_x$ and $\nu_y$ are generalized Poisson ratios, these read as
\begin{subequations}    
\begin{eqnarray} \label{eq:young}
        E &=& C_{11} - \frac{C_{11}C_{13}^2 + C_{33}C_{12}^2 - 2C_{12}C_{13}^2}{C_{11}C_{33} - C_{13}^2},\\
        \nu_x &=&  \frac{C_{33}C_{12} - C_{13}^2}{C_{11}C_{33} - C_{13}^2},\label{eq:nux}\\
        \nu_y &=& \frac{C_{11}C_{13} - C_{12}C_{13}}{C_{11}C_{33} - C_{13}^2}.\label{eq:nuy}
\end{eqnarray}
\end{subequations}

The displacement fields of the flexural modes are parallel with the axis $x$ and $y$, respectively, in first order in $qd$. The corresponding solutions read as
\begin{subequations}
    \begin{align} \label{eq:solFx}
        \mathbf{w}^{q}_{\textrm{F},x}(x,y) = A_{\textrm{F},x}(\hat{\mathbf{x}} - iqx\hat{\mathbf{z}}),\quad &\omega_{\textrm{F},x}(q) = \sqrt{\frac{EI_y}{\rho A}}q^2,\\
        \label{eq:solFy}
        \mathbf{w}^{q}_{\textrm{F},y}(x,y) = A_{\textrm{F},y}(\hat{\mathbf{y}} - iqy\hat{\mathbf{z}}), \quad &\omega_{\textrm{F},y}(q) = \sqrt{\frac{EI_x}{\rho A}}q^2,
    \end{align}
\end{subequations}
where $A_{\textrm{F},x}$ ($A_{\textrm{F},y}$) is the amplitude of the $x$ ($y$) mode, $A = a b$ is the area of the cross section, and  $I_x$ ($I_y$) is the moment of inertia around the $x$ ($y$) axis, which reads as
\begin{equation} \label{eq:momentofinertia}
    I_x = \int\limits_{-b/2}^{b/2}\int\limits_{-a/2}^{a/2} y^2 \dd x \dd y, \quad I_y = \int\limits_{-b/2}^{b/2}\int\limits_{-a/2}^{a/2} x^2 \dd x \dd y.
\end{equation}

Finally, the solution for the torsional wave, in first order in $qd$, reads as
\begin{subequations}
    \begin{align} \label{eq:solT}
        \mathbf{w}^{q}_\textrm{T}(x,y) = \phi_0 \left( \hat{\mathbf{z}} \cross \boldsymbol{\varrho} + iq\Phi(x,y)\hat{\mathbf{z}}
        \right),\\
        \omega_\textrm{T}(q) =  \sqrt{\frac{C}{\rho I}} q,
    \end{align}
\end{subequations}
where $\phi_0$ is the amplitude, $\boldsymbol{\varrho} = (x,\,y,\,0)^T$, $\Phi$ is the warping function, $I = I_x + I_y$, and $C$ is the torsional rigidity. The warping function does not induce any rotation of the cross-section, therefore it does not couple to the spin qubit via the lattice reorientation mechanism. For more details on the coupling mechanism, see Sec.~\ref{sec:relaxation}, and about the warping function, see App.~\ref{subsec:warping}. The torsional rigidity is calculated from the solution of the Poisson equation
\begin{equation}
    \Delta \chi = -1,
\end{equation} 
with the boundary conditions read as
\begin{equation}\label{eq:chiboundary}
    \chi = 0, \quad \textrm{if} \quad\left\{\begin{array}{l}
        x = \pm \frac{\tilde{a}}{2}\quad \textrm{or} \\
         y = \pm \frac{\tilde{b}}{2}, 
    \end{array}\right.
\end{equation}
where $\chi$ is defined on a distorted cross-section with sides 
\begin{equation}
    \tilde{a} = \sqrt[4]{\frac{C_{44}}{C_{66}}}a,\quad \tilde{b} = \sqrt[4]{\frac{C_{66}}{C_{44}}}b.
\end{equation}
Then, the torsional rigidity reads as 
\begin{equation}
    C = 4\sqrt{C_{44}C_{66}}\int\limits_{-\tilde{a}/2}^{\tilde{a}/2}\int\limits_{-\tilde{b}/2}^{\tilde{b}/2} \chi \dd y \dd x.
\end{equation}
For a detailed description and derivation, see App.~\ref{app:thin_rod}.

\begin{figure}[b]
    \centering
    \includegraphics[width=\linewidth]{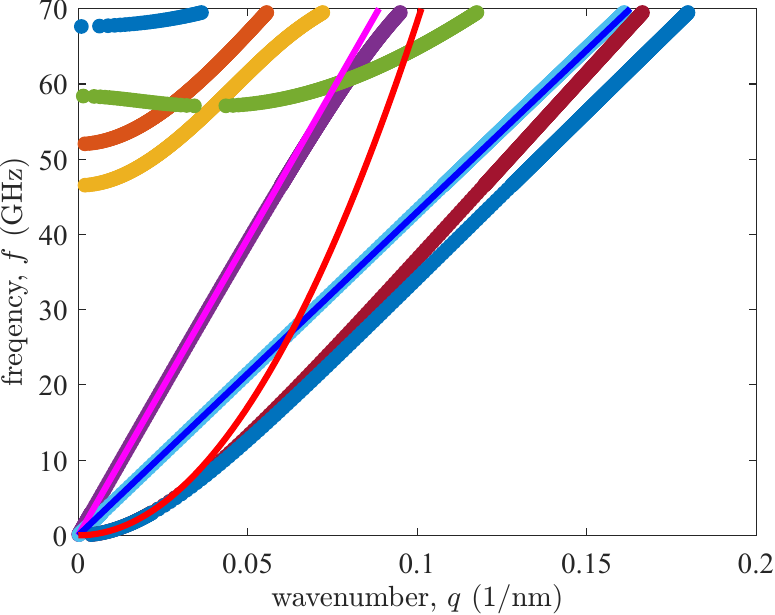}
    \caption{{Dispersion relation of phonons in a hexagonal germanium nanowire with square cross section.} Width of the nanowire is set to be $a = b = 30$ nm. 
    Thin solid lines: analytical, approximate dispersion relations obtained in the long-wavelength limit for longitudinal (purple),  torsional (blue) and flexural (red) modes. 
    Circles (forming thick solid lines): numerical results.
    The raw data  and the source code corresponding to this figure can be found in a dedicated GitHub repository \cite{gitrepo}.}
    \label{fig:dispersion}
\end{figure}

\subsection{Numerical results} \label{subsec:numerical_phonons}

We used the COMSOL Multiphysics software to numerically solve Eq.~\eqref{eq:reducedChristoffel} with free boundary conditions via the finite element method. 
For the numerical calculation, we assumed a square-shaped cross section, with width $a=b=30$ nm, and we used the values of stiffness tensor elements from Ref.~\cite{Mayengbam2023TheoreticalEngineering}.
We compare the numerical results with the analytical approximate results derived in the previous section in Fig.~\ref{fig:dispersion}.
This comparison shows agreement for the long-wavelength limit, establishing the credibility of the numerical results in the considered parameter range. Note that the two flexural modes are degenerate for sufficiently small wavenumbers, as expected from the analytical description when $I_y = I_x$, see Eqs.~\eqref{eq:solFx} and \eqref{eq:solFy}. This also implies that for a generic rectangular cross-section with $a\ne b$ the degeneracy would be absent.

The data generated by the COMSOL project can be found in a
dedicated GitHub repository \cite{gitrepo}.

\section{Relaxation of hole spin qubits via the lattice reorientation mechanism} \label{sec:relaxation}

In this section, we describe a coupling mechanism between a hole spin qubit defined in a nanowire quantum dot, and the acoustic phonons of the nanowire. 
This coupling mechanism is a consequence of the anisotropic bulk $g$ tensor, 
which is the consequence of spin-orbit interaction in a low-symmetry (hexagonal) crystal environment. 
In materials with higher symmetry, such as cubic GaAs, Si, or Ge, the spin-orbit interaction is present but leads to an approximately isotropic bulk $g$ tensor, and the coupling mechanism discussed here is therefore absent or strongly suppressed.
The physical origin of the coupling is that, in the presence of torsional or flexural phonons, the nanowire, and hence the underlying crystal lattice, undergoes local rotations. 
Since the $g$ tensor is tied to the crystal lattice through spin-orbit interaction, these rotations induce a reorientation of the $g$ tensor. 

The effect of $g$ tensor modulation due to strain fields has already been discussed in the literature \cite{roth1960gfactorrelaxation, glavin2003Spinlatticerelaxation}. 
Here, however, we consider $g$ tensor modulation arising from the rotational component of the displacement field, a mechanism that has also been discussed in the context of carbon nanotubes \cite{Borysenko2008ElectronNanotubes, Rudner2010SpinDots, Flensberg2010BendsCoupling, Palyi2012Spin-Orbit-InducedResonator}.
In the presence of a nonzero magnetic field, the rotation of the $g$ tensor can imply an effective transverse Zeeman field for the spin qubit, which leads to qubit relaxation. 
We call this the \textit{lattice reorientation} relaxation mechanism.

It should be noted that, even in highly symmetric crystals with isotropic bulk $g$ factor, a low-symmetry confinement in a nanostructure can lead to an anisotropic $g$ tensor, activating the lattice reorientation relaxation mechanism.

\subsection{General description of the lattice reorientation mechanism}
In this subsection, we derive the effective Hamiltonian of the hole spin qubit coupled to a phonon bath, using a phenomenological model based on the $g$ tensor formalism. 
Furthermore, we calculate the corresponding qubit relaxation rate from Fermi's Golden rule. 

To describe the spin-phonon coupling mechanism for the hole spin qubit, our tool is an effective, two level spin-qubit model which reads:
\begin{equation}\label{eq:Hfull}
    H = H_\textrm{q} + H_\textrm{int}.
\end{equation}
Here $H_\textrm{q}$ is the qubit Hamiltonian in the absence of deformation and $H_\textrm{int}$ is the interaction between the qubit and the phonons. 

The qubit Hamiltonian in the $g$ tensor formalism reads as
\begin{equation} \label{eq:Hq}
     H_\text{q} = \frac{1}{2} \muB \boldsymbol{\sigma} \cdot \mathbf{g}_0\mathbf{B},
\end{equation}
where $\muB$ is the Bohr magneton, and $\mathbf{g}_0$ is the material-specific bare $g$ tensor. 
In our case, we take the bulk $g$ tensor with the symmetries characteristic of hexagonal materials as the bare $g$ tensor:
\begin{equation}
\label{eq:bulkg0}
    \mathbf{g}_0 = \begin{pmatrix}
        g_\perp & 0 & 0 \\
        0 & g_\parallel & 0 \\
        0 & 0 & g_\perp 
    \end{pmatrix},
\end{equation}
with the numerical values describing hexagonal Ge, quoted in Table \ref{tab:bandparamaters}. Note that in the Eq.~\eqref{eq:bulkg0}, the $y$ axis corresponds to the $c$ axis of the crystal, as also indicated in Fig.~\ref{fig:nw_geom}.

The interaction term in the effective qubit model reads as:
\begin{equation}\label{eq:Hint}
    H_\textrm{int} = \frac{1}{2} \muB \boldsymbol{\sigma} \cdot \delta \mathbf{g}[\mathbf{u}]\mathbf{B},
\end{equation}
where $\delta \mathbf{g}[\mathbf{u}]$ is a functional of the displacement field $\mathbf{u}(\mathbf{r})$ describing the $g$ tensor modulation due to the reorientation of the lattice induced by the phonon displacement. 
We will express the interaction term up to linear order in the displacement $\mathbf{u}(\mathbf{r})$, implying that the interaction term will be decomposed as a sum of the contributions of the phonon modes. 
As we show below in detail, after this linearization, decomposition, and the quantization of the displacement field, we find:
\begin{equation} \label{eq:Hint_decomp}
    H_\textrm{int} = \sum_{\lambda, q}\frac{1}{2} \muB \boldsymbol{\sigma} \cdot \delta \mathbf{g}_\lambda^{q}\mathbf{B}(a_\lambda(q) + a_\lambda^\dagger(-q)),
\end{equation}
where $\delta \mathbf{g}_\lambda^{q}$ is a $3\times 3$ complex matrix, representing the contribution to the $g$ tensor modulation by the phonon mode with polarization index $\lambda$,
wavenumber $q$, and
annihilation operator $a_\lambda(q)$.

The mode-resolved $g$ tensor modulation matrix $\delta \mathbf{g}_\lambda^q$ depends on shape of the modes as well as the orbital envelope of the spin qubit.
In the next part of this section, we specify the orbital envelope and express the $\delta \mathbf{g}_\lambda^q$ contributions, following these steps:

(1) We specify a real-space, three-dimensional model of a spin-$\frac{1}{2}$ hole confined in a quantum dot in the absence of the magnetic field, as well as its ground-state orbital. 

(2) We formulate a phenomenological model of the lattice reorientation mechanism, i.e., model the modulation of the Zeeman term due to the displacement field in linear order.
    
(3) We project the displacement-field-dependent Zeeman term to the ground-state Kramers pair of the real-space, microscopic model of step 1. 
    In this step, we identify the $g$ tensor modulation functional $\delta\mathbf{g}[\mathbf{u}(\mathbf{r})]$, introduced in Eq.~\eqref{eq:Hint}.
    \label{step:projection}

(4) We quantize the displacement field to arrive to the form of Eq.~\eqref{eq:Hint_decomp}, and obtain an analytic expression for $\delta \mathbf{g}_\lambda^q$.

\textit{Step 1.} The Hamiltonian of the real-space effective-mass model of the quantum dot reads as
\begin{equation}
    H_\mu =  \frac{p_x^2 + p_z^2}{2m_\perp} + \frac{p_y^2}{2m_\parallel}  + V_\textrm{hw} (x,y) + \frac{\hbar^2}{2m_\perp\ld^4} z^2,
\end{equation}
where the first two terms form the kinetic energy, and last two terms describe the confinement of the hole. 
The parameters $m_\perp$ and $m_\parallel$ are the effective masses; we used the numerical values for hexagonal Ge, listed in Table \ref{tab:bandparamaters}. 

The transverse potential, $V_\textrm{hw} (x,y)$, is a hard-wall, infinite potential well, which reads as
\begin{equation}
    V_\textrm{hw} (x,y) = \left\{\begin{array}{lc}
       0  & \textrm{if}\,\left.\begin{array}{l}
            x \in [-a/2, a/2], \\
            y \in [-b/2, b/2], 
       \end{array}\right.   \\
        \infty & \textrm{else.} 
    \end{array} \right.
\end{equation}
The longitudinal confinement is modeled as a harmonic confinement, with ground-state broadening $\ld$.

The ground-state envelope function of the effective-mass Hamiltonian reads as
\begin{equation}\label{eq:groundmicro}
    \psi_\sigma(\mathbf{r}) = \frac{2}{\sqrt[4]{a^2b^2\ld^2\pi}} \cos\left(\frac{x\pi}{a}\right) \cos\left(\frac{y\pi}{b}\right)e^{-\frac{z^2}{2\ld^2}}\ket{\sigma},
\end{equation}
where $\sigma$ is a pseudospin index with values $\uparrow$ and $\downarrow$. 
This ground state is doubly degenerate at zero magnetic field, forming a Kramers pair therefore the states $\ket{\sigma}$ can be arbitrarily chosen. We fix the gauge such that for zero displacement the projection of the Zeeman term on the envelope functions $\Psi_\sigma$ results in Eq.~\eqref{eq:Hq} with the $g$ tensor defined in Eq.~\eqref{eq:bulkg0}.

\textit{Step 2.} Due to phonons, the nanowire may be locally rotated. We model this as the rotation of the $g$ tensor, the orientation of which is fixed to the lattice. This results in a local, displacement field dependent $g$ tensor modulation. The Zeeman term, with the modulated $g$ tensor, reads as
\begin{equation}\label{eq:zeemanmicro}
    H_\textrm{Zeeman} = \frac{1}{2}\muB \boldsymbol{\sigma}\cdot \mathbf{g}(\mathbf{u}(\mathbf{r}))\mathbf{B},
\end{equation}
where $\mathbf{g}(\mathbf{u}(\mathbf{r}))$ is the local, displacement field dependent $g$ tensor. Note that in case of zero displacement it is equal to the bare $g$ tensor, i.e.
\begin{equation}
    \mathbf{g}(\mathbf{u}(\mathbf{r})=\boldsymbol{0}) = \mathbf{g}_0.
\end{equation}

The modulation of the $g$ tensor, due to the reorientation of the lattice, reads as
\begin{align} \label{eq:gmodulation}
   \mathbf{g}(\mathbf{u}(\mathbf{r})) = \mathbf{R}(\mathbf{u}(\mathbf{r})) \mathbf{g}_0\mathbf{R}^T(\mathbf{u}(\mathbf{r})),
\end{align}
where the rotation tensors, expressed with the displacement field in linear approximation, read as
\begin{subequations}
\begin{align} \label{eq:infrottensor}
    \mathbf{R}(\mathbf{u}(\mathbf{r})) &= I + \mathbf{M}(\mathbf{u}(\mathbf{r})), \\
    M_{ij}(\mathbf{u}(\mathbf{r})) &= \frac{1}{2}\left(\frac{\partial u_i(\mathbf{r})}{\partial r_j} - \frac{\partial u_j(\mathbf{r})}{\partial r_i}\right). \label{eq:defM}
\end{align}
\end{subequations}
Substituting the infinitesimal rotation tensor in Eq.~\eqref{eq:infrottensor} into Eq.~\eqref{eq:gmodulation}, the $g$ tensor reads as
\begin{align} \label{eq:gmodulation_withM}
   \mathbf{g}(\mathbf{u}(\mathbf{r})) = \mathbf{g}_0 + [\mathbf{M}(\mathbf{u}(\mathbf{r})),\mathbf{g}_0],
\end{align}
where $[A,B]$ is the commutator of the matrices $A$ and $B$, and we used that the $3\times 3$ complex matrix $\mathbf{M}$ is anti-symmetric.

\textit{Step 3.} In order to obtain the effective, two-level Hamiltonian in Eq.~\eqref{eq:Hfull}, we project the displacement modulated Zeeman term, defined in Eq.~\eqref{eq:zeemanmicro}, to the ground state of the quantum dot, which is given in Eq.~\eqref{eq:groundmicro}. Substituting Eq.~\eqref{eq:gmodulation_withM} into Eq.~\eqref{eq:zeemanmicro}, one can separate the Hamiltonian into two parts. After the projection, the separated two terms are the ones in Eq.~\eqref{eq:Hfull}: a qubit part, same as in Eq.~\eqref{eq:Hq}, and an interaction part as it is in Eq.~\eqref{eq:Hint}, with the $g$ tensor modulation functional defined as
\begin{equation} \label{eq:dg_fromM}
    \delta \mathbf{g}[\mathbf{u}] = [\langle\mathbf{M} \rangle,\mathbf{g}_0],
\end{equation}
where $\langle . \rangle$ means the expectation value with respect the ground-state envelope of the quantum dot in Eq.~\eqref{eq:groundmicro}, i.e.
\begin{multline}
    \langle f \rangle = \frac{4}{ab\ld\sqrt{\pi}}\int\limits_{-\infty}^{\infty}\int\limits_{-\frac{b}{2}}^{\frac{b}{2}}\int\limits_{-\frac{a}{2}}^{\frac{a}{2}}f(\mathbf{r})\\
    \times\cos^2\left(\frac{x\pi}{a}\right) \cos^2\left(\frac{y\pi}{b}\right)e^{-\frac{z^2}{\ld^2}}\dd x \dd y \dd z.
\end{multline}

\textit{Step 4.} The quantization of the displacement field of a quasi one-dimensional nanowire yields the following result, see Appendix \ref{app:phononQuantization}: 
\begin{equation}\label{eq:displacement}
    \mathbf{u}(\mathbf{r}) = \sum_{\lambda, q}  \sqrt{\frac{\hbar}{2LA\rho \omega_\lambda (q)}}\mathbf{e}_\lambda^q (x,y) \left( a_\lambda (q) + a^\dagger_\lambda (-q)\right) e^{iqz},
\end{equation}
where $\mathbf{e}_\lambda^q (x,y)$ is the shape of the mode, with the normalization
\begin{equation} \label{eq:normalization}
    \int\limits_{-b/2}^{b/2}\int\limits_{-a/2}^{a/2}|\mathbf{e}_\lambda^q (x,y)|^2 \dd x \dd y = ab.
\end{equation}
Substituting the quantized displacement field into Eq.~\eqref{eq:defM}, we obtain the form of the infinitesimal rotation tensor as
\begin{equation} \label{eq:Mdecomp}
    \mathbf{M} = \sum_{\lambda,q} \mathbf{M}_{\lambda}^q\left( a_\lambda (q) + a^\dagger_\lambda (-q)\right),
\end{equation}
with the definition
\begin{equation}\label{eq:M_mode}
   M_{\lambda,ij}^q = \sqrt{\frac{\hbar}{8LA\rho \omega_\lambda (q)}} \left\langle\frac{\partial (e^\lambda_{q,i}e^{iqz})}{\partial r_j} - \frac{\partial (e^\lambda_{q,j}e^{iqz})}{\partial r_i}\right\rangle.
\end{equation}
The $g$ tensor modulation of an individual mode $\delta \mathbf{g}_\lambda^q$, in Eq.~\eqref{eq:Hint}, is then obtained by the substitution of Eq.~\eqref{eq:Mdecomp} into Eq.~\eqref{eq:dg_fromM}. It reads as
\begin{equation}\label{eq:dg_mode}
    \delta \mathbf{g}_\lambda^q = [\mathbf{M}_{\lambda}^q, \mathbf{g}_0].
\end{equation}

The relaxation rate due to the lattice reorientation mechanism, obtained via Fermi's Golden rule \cite{Dirac1927TheRadiation}, reads as
\begin{align}
    \Gamma &= \frac{2\pi}{\hbar}  \sum_{q_\textrm{f}, \lambda} |\bra{\textrm{e}, q_\textrm{f},\lambda} H_\textrm{int} \ket{\textrm{g},\textrm{vac}}|^2 \delta(\hbar \omega_{\lambda} (q_\textrm{f}) - \hbar\omegaq),
\end{align}
where $\ket{\textrm{g}}$ ($\ket{\textrm{e}}$) is the ground (excited) state of the qubit Hamiltonian in the absence of deformation, see Eq.~\eqref{eq:Hq}.
After substituting the form of the interaction from Eq.~\eqref{eq:Hint_decomp}, and transforming the summation into an energy integral, the relaxation rate reads as
\begin{multline}
    \Gamma =\frac{\pi\muB^2}{2\hbar }\sum_{(\lambda,q^*)}  \frac{2L}{h\left| v_\lambda(q^*)\right|} 
    \left| \mathbf{n} \cross \delta \mathbf{g}_{\lambda}^{q^*}\mathbf{B}\right|^2
    \\
    =\frac{L}{2} \left(\frac{\omegaq}{g^*}\right)^2\sum_{(\lambda,q^*)} \frac{1}{\left| v_\lambda(q^*) \right|}\left| \mathbf{n} \cross \delta \mathbf{g}_{\lambda^*}^{q^*}\mathbf{b}\right|^2. \label{eq:gamma}
\end{multline}
Here, the sum goes through all phonon modes with polarization index $\lambda$ and positive wavenumber $q^{*}>0$, for which $\omega_{\lambda^{*}} (q^*) = \omegaq$. We have also introduced the notations $\mathbf{b} = \mathbf{B}/B$, $g^* = |\mathbf{g}_0\mathbf{b}|$, $\mathbf{n} = \mathbf{g}_0\mathbf{b}/g^*$  for convenience. The function $v_\lambda(q^*)$ is the group velocity, defined as:
\begin{equation}\label{eq:groupvelocity}
    v_{\lambda} (q^*) = \left.\frac{\dd \omega_\lambda (q) }{\dd q}\right|_{q^*}
\end{equation}
The factor of two in the numerator of the summand in the first line of Eq.~\eqref{eq:gamma} is a consequence of the $q \leftrightarrow -q$ symmetry of the phonon modes. 

\subsection{Analytical evaluation of the spin relaxation rate in the limit of low Larmor frequency}\label{subsec:analytical_relaxation}

So far, we have expressed the spin-qubit relaxation rate via the frequencies and the shapes of the modes.
In the limit of weak magnetic field and low Larmor frequency, only those modes can contribute to relaxation, which have vanishing frequency for $q = 0$.
We have expressed the dispersion relations and shapes of those modes in Sec.~\ref{subsec:analdisp}.
Here, we utilize those results to obtain analytical formulas for the relaxation rate in Eq.~\eqref{eq:gamma}.
The lattice reorientation mechanism couples only the flexural and torsional modes to the spin; here, we descibe the effect of those modes. 
The longitudinal mode does not rotate the nanowire, hence it does not contribute to relaxation via this mechanism. 

\textit{Torsional mode.} The shape of the low-frequency torsional mode is obtained via the normalization of Eq.~\eqref{eq:solT} according to Eq.~\eqref{eq:normalization}. In leading order, the shape of the mode reads as
\begin{equation}
    \label{eq:eqT}
    \mathbf{e}_\textrm{T}^q(x,y) = \sqrt{\frac{A}{I}} \left(\hat{\mathbf{z}} \cross \boldsymbol{\varrho} + iq\Phi(x,y)\hat{\mathbf{z}}\right).
\end{equation}
Inserting Eq.~\eqref{eq:eqT} into Eq.~\eqref{eq:M_mode} yields
\begin{equation}\label{eq:MT}
    \mathbf{M}_{\textrm{T}}^q = \sqrt{\frac{\hbar}{2IL\rho \omega_\textrm{T}(q)}}e^{-\frac{\ld^2q^2}{4}}
    \begin{pmatrix}
        0 & -1 & 0\\
        1 & 0 & 0 \\
        0 & 0 & 0
    \end{pmatrix}.
\end{equation}
Here, the contribution to the rotation induced by the warping function $\Phi$ is neglected, because this contribution is first order in the small parameter $qd$, whereas Eq.~\eqref{eq:MT} is zeroth-order. Note that for a rectangular cross-section this contribution is zero due to the symmetry of the cross-section. The mirroring symmetry to the $x$ and $y$ axes imply the same symmetry of the ground-state envelope function. They also imply the following properties of the derivatives of the warping function:
\begin{subequations}\label{eqs:warpingsymmetries_main}
\begin{align}
    \frac{\partial \Phi(x,y)}{\partial x} = -\frac{\partial \Phi(x,-y)}{\partial x},\\
   \frac{\partial \Phi(x,y)}{\partial y} = -\frac{\partial \Phi(-x,y)}{\partial y}.
\end{align}
\end{subequations}
For more details see Appendix~\ref{subsec:warping}. Therefore Eq.~\eqref{eq:M_mode} for the warping function gives zero.

The $g$ tensor modulation, which is obtained from the substitution of Eq.~\eqref{eq:MT} into Eq.~\eqref{eq:dg_mode}, reads as
\begin{equation} \label{eq:dgT}
    \delta\mathbf{g}_{\textrm{T}}^q = \sqrt{\frac{\hbar}{2IL\rho \omega_\textrm{T}(q)}}e^{-\frac{\ld^2q^2}{4}} \Delta g
    \begin{pmatrix}
        0 & 1 & 0\\
        1 & 0 & 0 \\
        0 & 0 & 0
    \end{pmatrix},
\end{equation}
where $\Delta g = g_\perp - g_\parallel$. The group velocity is independent of $q$, due to the linear dispersion relation, see Eq.~\eqref{eq:solT}. It reads as
\begin{equation}\label{eq:vT}
    v_\textrm{T} (q^*) = \sqrt{\frac{C}{\rho I}}.
\end{equation}
Finally, the relaxation rate, obtained from Eq.~\eqref{eq:gamma}, using Eqs.~\eqref{eq:dgT} and \eqref{eq:vT}, reads as
\begin{equation} \label{eq:gammaT}
    \Gamma_\textrm{T} = \frac{\hbar \omegaq}{4\sqrt{\rho I C}} \left(\frac{\Delta g}{g_\perp}\right)^2 A_\textrm{T}(\theta, \phi, \gamma) e^{-\frac{\rho I \omegaq^2\ld^2}{2C}},
\end{equation}
where we introduced the auxiliary function $A_\textrm{T}$, which reads as
\begin{equation}\label{eq:AT}
    A_\textrm{T}(\theta, \phi, \gamma) = \sin^2\theta\frac{\cos^2\theta + \sin^2\theta(\cos^2\phi - \gamma \sin^2\phi)^2}{[\cos^2\theta + \sin^2\theta(\cos^2\phi + \gamma^2\sin^2\phi)]^2},
\end{equation}
where $\gamma = g_\parallel/g_\perp$ and $\theta$ and $\phi$ are the polar and azimuthal angles corresponding to the coordinate system in Fig.~\ref{fig:nw_geom}.

The relaxation rate grows linearly with increasing magnetic field as long as $\omegaq \ll \omega_\textrm{T}(q = 2\pi/\ld)$. 
We explain this linear dependence via power counting as follows. 
On the one hand, the quantized displacement field has an explicit frequency dependence following $1/\sqrt{\omega}$, see Eq.~\eqref{eq:displacement}.
This frequency dependence is inherited by $\mathbf{M}^{q}_{\textrm{T}}$, see Eq.~\eqref{eq:MT}.
On the other hand, the interaction Hamiltonian $H_\mathrm{int}$ in Eq.~\eqref{eq:Hint} is proportional to the magnetic field, hence scales linearly with $\omega$.
Together with the $\omega$-independent density of states of the torsional phonons, these imply a relaxation rate that is proportional to $\omega$.

\textit{Flexural mode.}
The flexural waves also rotate the cross section around the $x$ or $y$ axes, depending on the mode. The rotation around the $y$ axis does not change the $g$ tensor; in other words, the infinitesimal rotation tensor corresponding to the $x$ directional flexural mode commutes with the bare $g$ tensor, i.e.~$[\mathbf{M}_{\textrm{F},x},\mathbf{g}_0] = 0$. 

The $y$ directional flexural mode, which rotates the nanowire around the $x$ axis, does induce relaxation due to the lattice reorientation mechanism. 
The shape of this mode is obtained by the normalization of the displacement field in Eq.~\eqref{eq:solFy}, according to Eq.~\eqref{eq:normalization}; the mode shape reads as 
\begin{equation}\label{eq:eqFy}
    \mathbf{e}^q_{\textrm{F},y} = \hat{\mathbf{y}} - iqy \hat{\mathbf{z}},
\end{equation}
in first order in the small parameter $qd$. The infinitesimal rotation tensor, which is the result of the substitution of Eq.~\eqref{eq:eqFy} into Eq.~\eqref{eq:M_mode}, reads as
\begin{equation} \label{eq:MFy}
    \mathbf{M}^q_{\textrm{F},y} = \sqrt{\frac{\hbar}{2A L\rho \omega_{\textrm{F},y}(q)}}e^{-\frac{\ld^2q^2}{4}} iq
    \begin{pmatrix}
        0 & 0 & 0\\
        0 & 0 & 1 \\
        0 & -1 & 0
    \end{pmatrix}.
\end{equation}
The $g$ tensor modulation, obtained by inserting Eq.~\eqref{eq:MFy} into Eq.~\eqref{eq:dg_mode}, reads as
\begin{equation}\label{eq:dgFy}
    \delta\mathbf{g}_{\textrm{F},y}^q = \sqrt{\frac{\hbar}{2AL\rho \omega_{\textrm{F},y}(q)}}e^{-\frac{\ld^2q^2}{4}} iq\Delta g
    \begin{pmatrix}
        0 & 0 & 0\\
        0 & 0 & 1 \\
        0 & 1 & 0
    \end{pmatrix}.
\end{equation}
The group velocity, which is calculated from the derivation of the dispersion relation in Eq.~\eqref{eq:solFy}, reads as
\begin{equation} \label{eq:vFy}
    v_{\textrm{F},y}(q^*) = 2\sqrt{\frac{EI_x}{\rho A}} q^* = 2\sqrt[4]{\frac{EI_x\omegaq^2}{\rho  A}}.
\end{equation}

Inserting Eqs.~\eqref{eq:dgFy} and \eqref{eq:vFy} into Eq.~\eqref{eq:gamma} yields the following relaxation rate due to the low-energy flexural phonon:
\begin{equation} \label{eq:gammaF}
    \Gamma_\textrm{F} = \frac{1}{8} \sqrt[4]{\frac{\hbar^4\omega_\textrm{L}^6}{\rho AE^3I_x^3}} \left(\frac{\Delta g}{g_\perp}\right)^2 A_\textrm{F}(\theta, \phi, \gamma)e^{-\sqrt{\frac{\rho A}{EI_x}}\frac{\omega_\textrm{L}\ld^2}{2}},
\end{equation}
where we introduced the auxiliary function $A_\textrm{F}$, which reads as
\begin{widetext}
\begin{equation}\label{eq:AF}
    A_\textrm{F}(\theta, \phi, \gamma) = \frac{\sin^2\theta\cos^2\phi(\cos^2\theta + \sin^2\theta\sin^2\phi) + (\cos^2\theta - \gamma \sin^2\theta\sin^2\phi)^2}{[\cos^2\theta + \sin^2\theta(\cos^2\phi + \gamma^2\sin^2\phi)]^2}.
\end{equation}
\end{widetext}

According to Eq.~\eqref{eq:gammaF}, 
the relaxation rate of the flexural mode grows with the power of $3/2$ of the frequency in the $\omegaq \ll \omega_{\textrm{F},y}(q=2\pi/\ld)$ regime. This can be understood via the following power-counting argument. The displacement field has a frequency dependence of $1/\sqrt{\omega}$, which is, due to the quadratic dispersion relation in Eq.~\eqref{eq:solFy}, proportional to $1/q$. This is inherited by the infinitesimal rotation tensor $\mathbf{M}_{\textrm{F},y}^{q}$ besides a linear dependence on $q$ coming from the $z$ derivative of the $y$ component and from the $y$ derivative of the $z$ component of the shape of the mode in Eq.~\eqref{eq:solFy}. The two factors cancel each other, leading to a frequency independent infinitesimal rotation tensor. Thus, the only frequency dependence of the interaction Hamiltonian in Eq.~\eqref{eq:Hint} is the result of the proportionality to the magnetic field. In addition, due to the quadratic dispersion relation in Eq.~\eqref{eq:solFy} the group velocity in Eq.~\eqref{eq:vFy} is proportional to $\sqrt{\omega}$. Then taking the square of the matrix element of the interaction Hamiltonian and dividing by the absolute value of the group velocity results in an overall dependence of $\omega$ to the power of $3/2$.

\subsection{Sweet-spot analysis: the dependence of the relaxation rates on the magnetic field direction}

In this section, we analyze the analytically derived formulas in Eqs.~\eqref{eq:gammaT} and \eqref{eq:gammaF}, comparing them with numerical results obtained by evaluating Eq.~\eqref{eq:gamma} using the numerical phonon modes from Sec.~\ref{subsec:numerical_phonons}. Specifically, we examine how the relaxation rates depend on the strength and direction of the magnetic field and numerically determine the relaxation sweet spot as a function of the Larmor frequency.  

\begin{figure*}
    \centering
    \includegraphics[width=\linewidth]{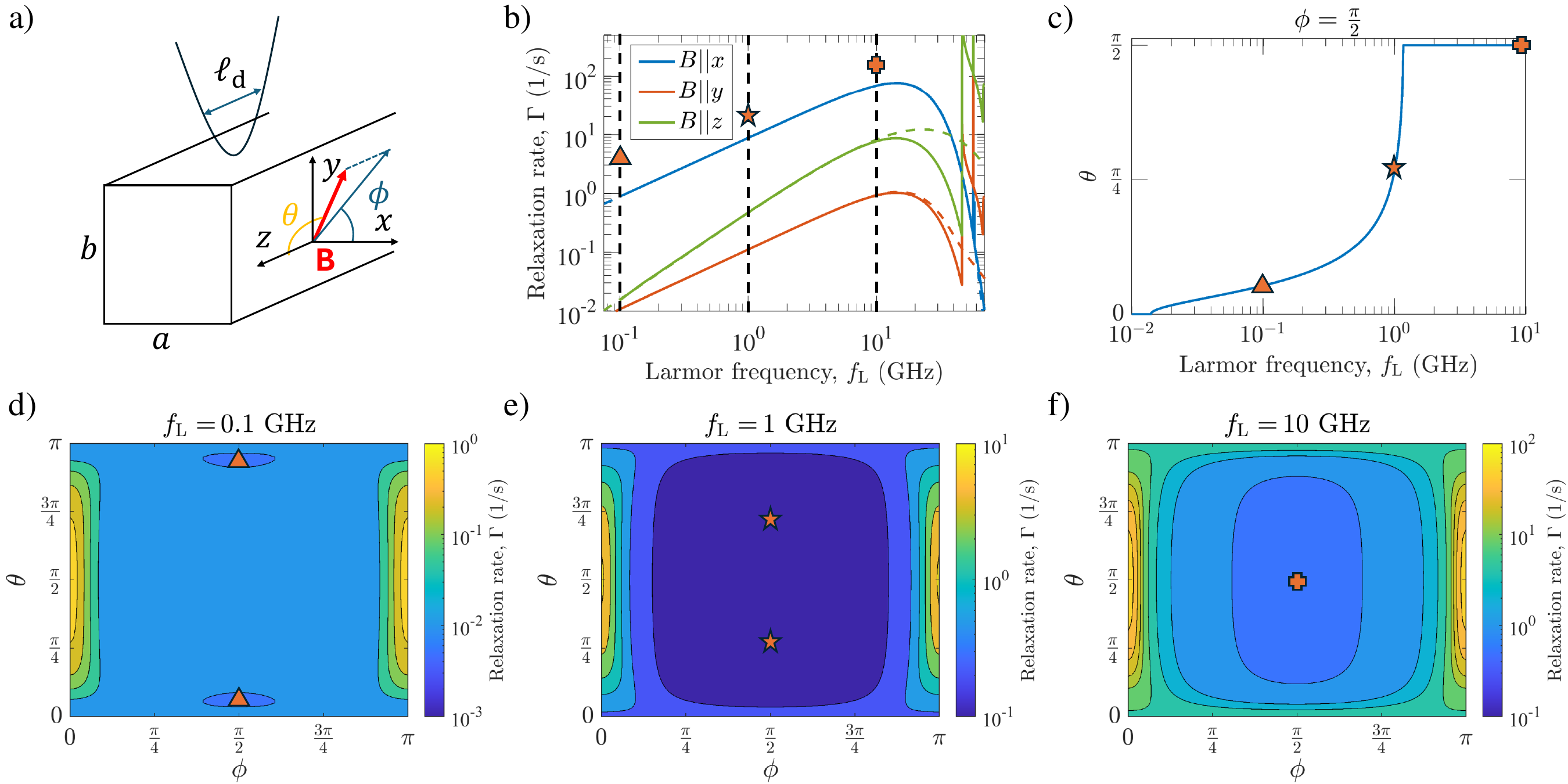} 
    \caption{{Relaxation rate of a hole spin qubit in a hexagonal germanium nanowire due to the lattice reorientation mechanism.} 
    (a) Angle parametrization of the magnetic field with respect to the nanowire. The orientation of the $c$ axis of the crystal points in the $y$ direction as it is in Fig.~\ref{fig:nw_geom}. 
    (b) Analytical results (dashed) for the spin qubit relaxation rate, based on Eq.~\eqref{eq:gammaT} and Eq.~\eqref{eq:gammaF}, for three special magnetic field directions. 
    Numerical results (solid) obtained by evaluating Eq.~\eqref{eq:gamma} using the numerical results for phonons from Sec.~\ref{subsec:numerical_phonons}. Numerical results exhibit van Hove singularities for $\fL \gtrsim 45$ GHz. The power-law dependence shows that for $B\,||\,z$ the torsional and for $B\,||\,x$ the flexural modes are decoupled from the qubit. Vertical dashed lines for the frequencies plotted in the d-e) panels are marked with triangle, star and cross, respectively.
    (c) Relaxation sweet spot as the function of the Larmor frequency. The sweet spot is in the $zy$ plane, along the magnetic field angle $\theta$ shown in the graph. Above ca.~1 GHz, relaxation sweet spot is at the $y$ direction, due to the large $g$ factor in that direction. Below ca.~1 GHz, the flexural mode contribution becomes negligible besides the torsional mode, therefore the sweet spot approaches the $z$ axis, where the torsional mode decouples from the qubit.
    [(d) and (e)] Relaxation rate dependence on the magnetic field orientation for different qubit frequencies. Markers indicate sweet spot. In (f), the sweet spot is in the $y$ direction, therefore there is no extra purple marker. As it is seen in panel (d), the sweet spot points close to the $z$ axis for $\fL = 0.1$ GHz. By carefully choosing the orientation of the magnetic field, the relaxation time can be kept above 1 second.
    The source code used to generate the (b)-(f) panels can be found in a
    dedicated GitHub repository \cite{gitrepo}.}
    \label{fig:gammamap}
\end{figure*}

Figure~\ref{fig:gammamap}(b) presents the numerical evaluation of Eq.~\eqref{eq:gamma} for $\Gamma$ (solid lines) alongside the analytical results from Sec.~\ref{subsec:analytical_relaxation} (dashed lines) for three special magnetic field directions. The calculations assume geometric parameters of $a = b = 30$ nm and $\ld = 30$ nm. Note that although the condition $a=b$ seems fine tuned, the qualitative behavior would not change if one resolves this condition because the symmetry group of the hex-Ge nanowire would not change. At low Larmor frequencies, the numerical and analytical results align closely. However, at higher frequencies, deviations occur, particularly at Larmor frequency values where the numerical results diverge. These singularities arise due to the van Hove singularities~\cite{vanHove1953} in the phonon density of states, which originate from gapped phonon modes that are not included in the analytical calculations, see Fig.~\ref{fig:dispersion}.  

A notable feature in the results is the difference in the power-law behavior at low frequencies. When the magnetic field is aligned with the $z$ axis, the power-law exponent is $3/2$, whereas for the other two cases, it is $1$. This difference arises because the exponent of $1$ is characteristic of the torsional mode, which rotates the nanowire around the $z$ axis. When the magnetic field is in the $z$ direction the torsional mode decouples from the spin qubit. Mathematically, this means that the $g$-tensor modulation in Eq.~\eqref{eq:dgT} maps the $\hat{\mathbf{z}}$ unit vector to zero. Consequently, for small Zeeman fields in the $z$ direction, only the $y$ directional flexural mode couples to the qubit, leading to the $3/2$ power law, as shown in \eqref{eq:gammaF}.  

Similarly, the $y$ directional flexural mode is decoupled for a magnetic field in the $x$ direction because it rotates the nanowire around the $x$ axis and the $\hat{\mathbf{x}}$ unit vector is in the kernel of the $g$-tensor modulation, see Eq.~\eqref{eq:dgFy}. This results in a power-law exponent of $1$ for that direction. When the magnetic field points in the $y$ direction, both modes contribute, and the overall power-law behavior is determined by the lower exponent, which is $1$. 

\begin{table}[]
    \caption{{Couplings between the spin qubit and different phonon modes for special magnetic-field directions.} The longitudinal and the $x$ directional flexural modes are always decoupled from the qubit in the context of lattice reorientation mechanism, because the longitudinal mode does not rotate the nanowire, while the $x$ directional flexural mode rotates around the $y$ axis, which is a symmetry of the bare $g$ tensor. The $y$ directional flexural mode rotates around the $x$ axis therefore it is decoupled from the qubit when the magnetic field is parallel to the $x$ axis. Similarly, the torsional mode rotates around the $z$ axis, therefore it is decoupled when the magnetic field points in the $z$ direction.}
    \label{tab:decoupling}
    \centering
    \begin{tabularx}{\linewidth}{lYYYc} \hline\hline
    \noalign{\vskip 4pt}
        \diaghead{\hskip35pt}{$\mathbf{b}$}{$\lambda$} & longitudinal & flexural $x$ & flexural $y$ & torsional \\[2pt] \hline
        \noalign{\vskip 3pt}
        $\mathbf{b} || x$ & off & off & off& on \\ 
        $\mathbf{b} || y$  & off & off & on & on \\ 
        $\mathbf{b} || z$  & off & off & on & off \\[2pt] \hline\hline
    \end{tabularx}

\end{table}

This analysis demonstrates that relaxation rates are highly sensitive to the magnetic field direction, with certain phonon modes decoupling under specific orientations. Table~\ref{tab:decoupling} summarizes these decoupling conditions. The dependence of relaxation rates on the magnetic field direction is influenced by two key factors: (1) the angle between axis of rotation induced by the phonon mode and the Zeeman field direction, and (2) the anisotropy of the bare $g$ tensor, which enters Eq.~\eqref{eq:gamma} through the $(1/g^*)^2$ factor. The $g$ tensor is maximal along the $y$ direction and minimal in the $xz$ plane.  

Identifying the relaxation sweet spot, i.e., the magnetic-field direction along which the relaxation rate is minimized for a fixed Larmor frequency, is nontrivial due to the interplay of these two effects. We found that the sweet spot lies in the $yz$ plane for the parameter set that we have considered. We plot the polar angle $\theta$ of the sweet-spot direction as a function of the Larmor frequency in Fig.~\ref{fig:gammamap}(c). At low frequencies, the torsional mode dominates due to its lower power-law exponent and is decoupled for fields along $z$, that is why the sweet spot tends to the $z$ axis.  For Larmor frequencies above $1$ GHz, the sweet spot aligns with the $y$ direction, where the $g$ factor is significantly larger than in the $xz$ plane, so for higher frequencies factor (2) becomes dominant. 

Finally, Fig.~\ref{fig:gammamap}(d)–\ref{fig:gammamap}(f) illustrates the relaxation rates for all magnetic field directions at three different Larmor frequencies. At low frequencies, the two anisotropic effects lead to a complex angular dependence. At higher frequencies, the anisotropy of the bare $g$ factor becomes the dominant factor, as the relaxation rates of different phonon polarizations become comparable.  

\section{Discussion} \label{sec:discussion}

In this section we discuss the limitations of our results and highlight further mechanisms that are expected to cause spin relaxation in hexagonal germanium nanowires.

\textit{Boundary conditions.} 
In our acoustic phonon calculations we used the simplest, floating boundary conditions, which is an established method in the literature \cite{Trif2008SpinLine, Bulaev2008Spin-orbitDots, Droth2013ElectronDots, Kloeffel2014AcousticNanowires}. However, in real devices, the nanowires are either suspended at one or both ends, or put on a substrate. 
The different boundary conditions change the dispersion relations and the mode shapes, which in turn change the relaxation rates as well. 
The methods we used here, and the results we obtained, offer a convenient starting point to analyze any of these scenarios. 
In particular, we expect the floating boundary condition to provide a good approximation when the nanowire is suspended and sufficiently long compared to the phonon wavelength associated with the qubit's Larmor frequency. For example, at a typical frequency of 1~GHz, the relevant wavelengths are approximately 520~nm for the flexural mode and 2.7~$\mu$m for the torsional mode. Therefore, our model yields quantitatively accurate predictions only when the nanowire length exceeds several microns and the Larmor frequency is above 1~GHz.

\textit{Further spin-phonon coupling mechanisms.} Here, we analyzed a spin-phonon coupling caused by the strong anisotropy of the bulk $g$ tensor. 
Another possible spin-phonon interaction mechanism involves deformation-potential (DP) coupling combined with Rashba-type spin-orbit interaction, which can arise from inversion symmetry breaking due to, for example, a substrate or gate-induced electric fields. 
Interband couplings due to mechanical strain may also contribute.
Although the DP parameters and the structure of Rashba-type spin-orbit interaction in hexagonal germanium have not yet been established experimentally or via \textit{ab initio} methods, we can still compare the expected Larmor-frequency dependence of the two mechanisms through a simple power-counting analysis.

For the longitudinal acoustic phonon mode, the matrix element associated with DP coupling scales as $\omega^{5/2}$: one power arises from the Van Vleck cancellation \cite{Khaetskii2000SpinDots}, another from the dipole approximation, a third from the gradient of the displacement field and a $1/\sqrt{\omega}$ factor comes from the quantized displacement field, see Eq.~\eqref{eq:displacement}. 
The constant (frequency-independent, $\propto \omega^0$) density of states of the linearly dispersing longitudinal mode does not contribute to the exponent, leading to the DP relaxation rate scaling as $\Gamma_{\textrm{DP,L}} \propto \omega^5$.

For flexural modes, the deformation tensor is quadratic in $q$, which implies a linear dependence on the Larmor frequency due to the quadratic dispersion relation, see Eq.~\eqref{eq:solFx}. 
Together  with the dipole approximation, the $1/\sqrt{\omega}$ factor in the quantized displacement field and the van Vleck cancellation, the matrix element depends quadratically on the Larmor frequency. 
The density of states adds an additional $1/\sqrt{\omega}$ factor, which results in an overall $\Gamma_{\textrm{DP,F}} \propto \omega^{7/2}$ dependence.

In contrast, the lattice reorientation mechanism has lower power dependence: for the torsional mode $\Gamma_{\textrm{LR,T}} \propto \omega$ and for the flexural mode $\Gamma_{\textrm{LR,F}} \propto \omega^{3/2}$. 
Therefore, at sufficiently low magnetic fields, one expects the reorientation mechanism to dominate over the deformation-potential-induced spin relaxation. However, due to the lack of material parameters, such as deformation-potential constants and spin-orbit interaction strength in hexagonal Ge nanowires, we cannot determine what “sufficiently low” means in practice. 
It is entirely possible that the crossover field lies in an unphysically low regime.

\textit{Further information loss mechanisms: charge and hyperfine noise.} The aim of our work is to analyze phonon-induced spin relaxation, and we have not included other information loss mechanisms, such as hyperfine interaction or charge noise in our model. A quantitative treatment of charge noise would require detailed knowledge of the device geometry, electrostatic environment, charge traps, and dielectric disorder, all of which are highly device-specific and beyond the scope of this manuscript. The analysis of hyperfine noise would require the hyperfine coupling in hexagonal germanium, which is not yet determined.
Note that relaxation due to phonons provides an upper bound on the qubit’s information-loss time scale. However, we do not claim to estimate coherence times, as several information loss mechanisms, such as charge noise, are excluded from our analysis.

\textit{Microscopic modeling of the reorientation mechanism.} While our approach is phenomenological, it could in principle be anchored in a microscopic theory. One possible direction would be to construct a simplified microscopic model in which the effect of a local lattice rotation on the $g$ tensor can be analyzed explicitly like in Refs.~\cite{Izumida2009SpinOrbitAnalysis, Rudner2010SpinDots, Flensberg2010BendsCoupling, Klinovaja2011CarbonFields, Ochoa2012Spin-orbitGraphene, Palyi2012Spin-Orbit-InducedResonator}. Alternatively, one could pursue a more realistic approach based on atomistic simulations, e.g., density functional theory (DFT) or DFT-derived tight-binding models of hexagonal Ge nanowires using methods from Refs.~\cite{Noffsinger2010EPW:Functions, Giustino2017Electron-phononPrinciples}. Such studies could provide quantitative input for the dependence of the $g$ tensor on strain and rotation of the nanostructure. Applying such techniques to hexagonal Ge nanowires would be an interesting direction for future work, but is beyond the scope of the present manuscript. 

\section{Conclusion} \label{sec:conclusion}

We have analyzed the geometric constraints of efficient electron or hole confinement, the acoustic phonons, the lattice reorientation spin-phonon coupling mechanism for holes, and relaxation of hole spin qubits, in hexagonal germanium nanowires.

We established the feasibility of electrostatically confined quantum dots in hexagonal germanium nanowires by identifying key geometric constraints necessary for optimal qubit operation. 

We computed the acoustic phonon modes using both analytical approximations and numerical methods. The numerical results validated our analytical approximations in the long-wavelength limit, reinforcing the reliability of our theoretical predictions and providing a solid foundation for further studies exploring the effects of phonons and electron-phonon interaction in hexagonal germanium nanowires.

We examined the impact of the lattice reorientation mechanism on the relaxation rate of a hole spin qubit. By deriving analytical expressions for qubit relaxation time, we demonstrated that relaxation rates depend significantly on both the amplitude and the direction of the applied magnetic field. The presence of relaxation sweet spots highlights a potential strategy for mitigating decoherence in hexagonal germanium-based qubits. Our results indicate that by optimizing the magnetic field orientation, relaxation rates can be effectively suppressed, enabling qubit lifetimes that exceed one second under appropriate conditions.
Remarkably, we find that the optimal magnetic field orientation, yielding the longest qubit relaxation time, depends on the qubit's Larmor frequency.
For the mechanism and parameter range we described, we found relaxation rates of at least 10 ms. 

These insights contribute to the broader effort of developing spin-based quantum technologies by providing important design principles for qubit implementations in hexagonal germanium nanowires. While our study focused on spin relaxation via the lattice reorientation mechanism, other spin-phonon coupling pathways---such as those arising from deformation potential electron-phonon coupling and spin-orbit interaction---may also play a role in qubit performance. Future research should aim to quantify these additional mechanisms and assess their relative impact in realistic device architectures.

\begin{acknowledgments}
The authors thank A.~Kossa, G.~Sz\'echenyi, and D.~Varjas for useful discussions.
This research was supported by the Ministry of Culture and Innovation (KIM) and the National Research, Development and Innovation Office (NKFIH) within the Quantum Information National Laboratory of Hungary (Grant No.~2022-2.1.1-NL-2022-00004), by the HUN-REN Hungarian Research Network through the HUN-REN-BME-BCE Quantum Technology Research Group, and by the European Union within the Horizon Europe research and innovation programme via the ONCHIPS project under grant agreement No 101080022.
The project supported by the Doctoral Excellence Fellowship Programme (DCEP) is funded by the National Research Development and Innovation Fund of the KIM and the Budapest University of Technology and Economics, under a grant agreement with the NKFIH.

B.K.~was the major contributor to all research aspects of this work, including the writing of the manuscript. 
Gy.F.~contributed to the theoretical investigation of nanowire phonons.
A.P.~acquired funding, supervised the project, and reviewed and edited the manuscript.
\end{acknowledgments}

\section*{Data availability}
The data that support the findings of this article are openly
available \cite{gitrepo}.

\appendix
\section{Acoustic phonon dispersion relation and displacement field in rectangular nanowires in the long-wavelength limit} \label{app:thin_rod}

In this appendix, we derive the equations of motion of a nanowire with rectangular cross-section.
We focus on a nanowire with hexagonal symmetry, as described in the main text, and describe the long-wavelength limit here.
The long-wavelength limit is the case when the displacement field varies slowly compared to the diameter of the cross section $d = \max(a,b)$, where $a$ and $b$ are the side lengths of the cross section.
Formally, we consider the dimensionless product $qd$, where $q$ is the wavenumber in the longitudinal direction, as a perturbative parameter, $qd \ll 1$.

In this section, we first state our assumptions. Then we use these assumptions, together with the translational invariance in the $z$ direction, to describe the displacement field with 7 generalized coordinates, which do not depend on the position. We express the Lagrangian of the system with these 7 coordinates, derive their equations of motion, and solve them, perturbatively in $q$. 
Finally, we discuss the solutions, and check that they are consistent with the assumptions we made to derive them.

\subsection{New coordinates and assumptions for the displacement field} \label{subsec:assumptions}

In this section, we formulate the assumptions in a form that holds for thin rods, we introduce 7 generalized coordinates that do not depend on the position, and express the stress and deformation tensors with these coordinates. 
Our assumptions are generally used in the literature of beam theory \cite{Vlasov1961Thin-walledBeams, Landau1986TheoryElasticity, Gurtin1973, Davi1996DynamicsRods}.
We will discuss their applicability in Appendix~\ref{subsec:assumptions_discussion}

Due to the translational invariance in the $z$ direction, the $z$ dependence can be factorized as
\begin{equation}
    \mathbf{u}(\mathbf{r}, t) = \sum_{q}\tilde{\mathbf{w}}^q(x,y,t) e^{iqz}.
\end{equation}

We introduce 6 generalized coordinates for each wavenumber $q$. These read as:
\begin{subequations}
\begin{align}
\mathbf{v}^q &= \frac{1}{ab} \int\limits_{-b/2}^{b/2}\int\limits_{-a/2}^{a/2} \tilde{\mathbf{w}}^q \dd x \dd y,\\
\phi^q &= \frac{1}{2ab} \int\limits_{-b/2}^{b/2}\int\limits_{-a/2}^{a/2} \left(\frac{\partial \tilde{w}^q_y}{\partial x} - \frac{\partial \tilde{w}^q_x}{\partial y}\right) \dd x \dd y,\\
\vartheta_x^q &= \frac{1}{ab} \int\limits_{-b/2}^{b/2}\int\limits_{-a/2}^{a/2} \frac{\partial \tilde{w}^q_z}{\partial x} \dd x \dd y, \\
\vartheta_y^q &= \frac{1}{ab} \int\limits_{-b/2}^{b/2}\int\limits_{-a/2}^{a/2} \frac{\partial \tilde{w}^q_z}{\partial y} \dd x \dd y.
\end{align}
\end{subequations}
The vector $\mathbf{v}^q$ describes the average displacement of the cross section, $\phi^q$ can be interpreted as the rotation angle of the cross section around the $z$ axis and the coordinates $\vartheta^q_x$ and $\vartheta^q_y$ are angles, which describe the rotation of the cross section around the $y$ and $x$ axis respectively.

With these coordinates, $\tilde{\mathbf{w}}^q$ reads as
\begin{equation} \label{eq:rod_displacement}
    \tilde{\mathbf{w}}^q(x,y,t) = \mathbf{v}^q(t) + \phi^q(t) \hat{\mathbf{z}} \cross \boldsymbol{\varrho} + (\boldsymbol{\vartheta}^q(t) \cdot \boldsymbol{\varrho})\hat{\mathbf{z}} + \boldsymbol{\Psi}^q (x,y,t),
\end{equation}
where $\boldsymbol{\varrho} = (x,\,y,\,0)^T$.
Here, $\boldsymbol{\Psi}^q$ is the remaining part of $\tilde{\mathbf{w}}^q$, which has the property that
\begin{equation}\label{eq:warping_zeroavg}
    \int\limits_{-b/2}^{b/2}\int\limits_{-a/2}^{a/2} \boldsymbol{\Psi}^q \dd x \dd y = \boldsymbol{0}.
\end{equation}

Our assumption 1 is that although $\Psi_z^q$ depends on $q$ and $t$ in general, the dependence can be separated as a scalar factor, i.e.~$\Psi_z^q$ reads as
\begin{equation} \label{eq:warping_factorize}
    \Psi_z^q(x,y,t) = \eta^q(t)\Phi(x,y),
\end{equation}
where $\eta^q$ is the seventh generalized coordinate and $\Phi$ is the warping function. 
The warping function only depends on the geometry of the nanowire, and can be calculated from a static deformation. 
Note that $\eta^q$ and $\Phi(x,y)$ are only defined up to a constant factor, we fix this factor and determine $\Phi$ in Appendix~\ref{subsec:warping}.

Assumption 2 is that the in-plane stress is zero not only at the boundaries, but at any point of the cross section, i.e.,
\begin{equation}\label{eq:thinstress}
    \sigma_{xx} = \sigma_{xy} = \sigma_{yy} = 0.
\end{equation}
This is reasonable due to the small thickness of the nanowire.

Assumption 3 is that the contributions of the components $x$ and $y$ of $\boldsymbol{\Psi}^q$ to the Lagrangian are negligible.

Due to assumption 2, the linear relation in Eq.~\eqref{eq:hooke} between the stress and deformation, using the form of the stiffness tensor in Eq.~\eqref{eq:Chex}, reads as
\begin{equation} \label{eq:stress_strain}
    \sigma_{zz} = E\epsilon_{zz},  \quad \sigma_{xz} = 2C_{66}\epsilon_{xz},\quad \sigma_{yz} = 2C_{44}\epsilon_{yz},
\end{equation}
where $E$ is Young's modulus defined in Eq.~\eqref{eq:young}, and Eqs.~\eqref{eq:thinstress} and \eqref{eq:stress_strain} also imply constraints on the $\boldsymbol{\Psi}^q$. These read as
\begin{subequations} \label{eqs:stress_constraint}
\begin{gather}
    \Psi^q_{x,x} = -\nu_x \epsilon_{zz},\quad \Psi^q_{y,y} = -\nu_y \epsilon_{zz},\\
    \Psi^q_{x,y} + \Psi^q_{y,x} = 0. 
\end{gather}
\end{subequations}
In the above equations, we introduced the generalized Poisson's ratios defined in Eqs.~\eqref{eq:nux} and \eqref{eq:nuy}.

We decompose the deformation and stress tensors as Fourier sums for the wavenumber: 
\begin{equation} \label{eq:straindecomp}
    \epsilon_{\alpha\beta}(\mathbf{r}) = \sum_q \epsilon^q_{\alpha\beta}(x,y) e^{iqz}, \quad \sigma_{\alpha\beta}(\mathbf{r}) = \sum_q \sigma^q_{\alpha\beta}(x,y) e^{iqz}.
\end{equation}
The relevant elements of the deformation tensor read as: 
\begin{subequations} \label{eqs:strainthin}
    \begin{eqnarray}
        \epsilon_{zz}^q &=& iq(v^q_{z} + \boldsymbol{\vartheta}^q \cdot \boldsymbol{\varrho} + \eta^q \Phi),\\
        2\epsilon_{xz}^q &=& \vartheta^q_{x} + \eta^q \Phi_{,x} + iq(v^q_{x} - y\phi^q),\\
        2\epsilon_{yz}^q &=& \vartheta^q_{y} + \eta^q \Phi_{,y} + iq(v^q_{y} + x\phi^q).
    \end{eqnarray}
\end{subequations}
Here, we used assumption 1 and assumption 3, and introduced the following notation for the partial derivatives:
\begin{equation}
    f_{,\alpha} = \frac{\partial f}{\partial r_{\alpha}}, \quad f_{,\alpha\beta} = \frac{\partial^2 f}{\partial r_{\alpha}\partial r_{\beta}},\quad \dots
\end{equation}

The form of the Lagrangian in any elastic system reads as
\begin{eqnarray}\label{eq:Lagrangian}
    \mathcal{L} = \frac{1}{2}\int\dd V \sum_{ij}\epsilon_{ij}\sigma_{ij} - \rho \Dot{\mathbf{u}}^2.
\end{eqnarray}
Substituting the form of the stress elements from Eq.~\eqref{eq:stress_strain} into Eq.~\eqref{eq:Lagrangian} we obtain the Lagrangian of the nanowire, which reads as
\begin{multline} \label{eq:thin_lagrangian}
    \mathcal{L} = \int\limits_0^L \int\limits_{-b/2}^{b/2}\int\limits_{-a/2}^{a/2} \left[E\frac{\epsilon^2_{zz}}{2} + 2C_{66}\epsilon^2_{xz} + 2C_{44}\epsilon^2_{yz} \right.\\ \left. 
    - \frac{\rho}{2} \Dot{\mathbf{u}}^2\right]\dd x \dd y \dd z.
\end{multline}
Due to assumptions 1 and 3, the time and position dependence of the displacement field is factorized, hence the spatial integrals of this Lagrangian can be evaluated.
But before substituting the form of the stress and deformation tensors from Eq.~\eqref{eqs:strainthin}, we analyze the properties of the warping function $\phi(x,y)$ and present a recipe for calculating it numerically.

\subsection{Calculation of the warping function and its symmetry properties} \label{subsec:warping}

In this section, we present a methodology for calculating the warping function $\Phi$. The phenomenon described by the warping function is the following.
Upon twisting of a rod, the cross section rotates around the $z$ axis, and to satisfy the boundary conditions (i.e., that $\sigma_{xz}$ and $\sigma_{yz}$ must vanish at the boundary facets perpendicular to the $x$ and $y$ axes, respectively), the cross section bends in the $z$ direction. 
This deformation results in the term $\eta^q \Phi \hat{\mathbf{z}}$ in Eq.~\eqref{eq:rod_displacement}, which appears explicitly by invoking Eq.~\eqref{eq:warping_factorize}.

Due to assumption 1, the warping function does not depend on $q$ and $t$, therefore it can be calculated from a specific setup: a uniform and static twisting. Under such deformation, the displacement field reads as
\begin{equation}\label{eq:utwist}
    \mathbf{u}(\mathbf{r}) = \phi(z) \hat{\mathbf{z}} \cross \boldsymbol{\varrho}  + \eta (z) \Phi (x,y) \hat{\mathbf{z}}, 
\end{equation}
and uniformity means that $\phi_{,z}$ is constant.
In pure twisting, there is no axial stress in the system, therefore $\eta$ is independent of $z$. Moreover, $\phi_{,z} = 0$ corresponds to zero twist, which means that for infinitesimal deformation $\eta = \phi_{,z}$, where we merged the proportionality factor into $\Phi$. The shear deformation, calculated from Eq.~\eqref{eq:utwist},
reads as
\begin{equation} \label{eq:twist_shear}
        2\epsilon_{xz} =  \phi_{,z}(\Phi_{,x} - y),\quad
        2\epsilon_{yz} = \phi_{,z}(\Phi_{,y} + x).
\end{equation}
The equation of equilibrium, i.e.~that there is no force acting in the system, reads as
\begin{equation} \label{eq:torsion_eq}
        \sigma_{xz,x} + \sigma_{yz,y} = 0.
\end{equation}

This essentially means that the vector field $(\sigma_{xz}, \sigma_{yz})^T$ has zero divergence. Therefore there exist a scalar function $\chi (x,y)$, such that
\begin{equation} \label{eq:chi}
    \sigma_{xz} = 2\mu\phi_{,z} \chi_{,y}, \quad \sigma_{yz} = -2\mu \phi_{,z} \chi_{,x},
\end{equation}
where $\mu = \sqrt{C_{44}C_{66}}$. From Eq.~\eqref{eq:twist_shear} the following relation holds:
\begin{equation}
\label{eq:compatibility}
    2\epsilon_{xz,y} - 2\epsilon_{yz,x} = -\phi_{,z}.
\end{equation}
Using the linear relation between the shear stress and strain, as in Eq.~\eqref{eq:stress_strain}, and Eq.~\eqref{eq:chi} one obtains from Eq.~\eqref{eq:compatibility} a differential equation for $\chi$ of the form:
\begin{equation}
\label{eq:pre-poisson}
    \sqrt{\frac{C_{66}}{C_{44}}}\chi_{,xx} + \sqrt{\frac{C_{44}}{C_{66}}}\chi_{,yy} = -1.
\end{equation}

The differential equation \eqref{eq:pre-poisson} is equivalent to the Poisson equation with the following coordinate transformation:
\begin{gather}
    \tilde x = \sqrt[4]{\frac{C_{44}}{C_{66}}}x,\,\,\, \tilde y = \sqrt[4]{\frac{C_{66}}{C_{44}}}y \implies \chi_{,\tilde{x}\tilde{x}} + \chi_{,\tilde{y}\tilde{y}} = -1.
\end{gather}
The boundary conditions of the Poisson equation follow from the original problem, stated in Eqs.~\eqref{eqs:boundary}. These imply that 
\begin{subequations} \label{eqs:torsional_boundary}
\begin{align}
    \chi_{,\tilde{y}} = 0\quad \text{at} \quad x = \pm \frac{\tilde{a}}{2},  \\
    \chi_{,\tilde{x}} = 0\quad \text{at} \quad y = \pm \frac{\tilde{b}}{2},
\end{align}
\end{subequations}
which is equivalent to the property, that $\chi$ is constant at the boundary and can be set to zero, which is stated in Eq.~\eqref{eq:chiboundary}. This boundary value problem can be solved numerically. Note that $\chi$ has the property
\begin{equation}\label{eq:chisymmetry}
    \chi(x,y) = \chi(-x,y) = \chi(x,-y)
\end{equation}
due to the symmetry of the Poisson equation and its boundary conditions. 

The partial derivatives of the warping function, from Eqs.~\eqref{eq:stress_strain} and \eqref{eq:chi}, read as
\begin{subequations}
\begin{align} 
    \Phi_{,x} = 2\sqrt{\frac{C_{44}}{C_{66}}} \chi_{,y} + y,\\
    \Phi_{,y} = -2\sqrt{\frac{C_{66}}{C_{44}}} \chi_{,x} - x,
\end{align}
\end{subequations}
from which the warping function is uniquely determined due to the condition in Eq.~\eqref{eq:warping_zeroavg}. 
Note that due to the symmetry of the auxiliary function $\chi$, see Eq.~\eqref{eq:chisymmetry}, the derivatives of the warping function have the symmetry
\begin{subequations}\label{eqs:warpingsymmetries}
\begin{align}
    \Phi_{,x}(x,y) = -\Phi_{,x}(x,-y),\\
    \Phi_{,y}(x,y) = -\Phi_{,y}(-x,y).
\end{align}
\end{subequations}
These symmetries imply that
\begin{subequations} \label{eqs:zero_integrals}
\begin{align}
    \int\limits_{-b/2}^{b/2}\int\limits_{-a/2}^{a/2} x\Phi \dd x \dd y = \int\limits_{-b/2}^{b/2}\int\limits_{-a/2}^{a/2} \Phi_{,x}\left(\frac{a^2}{8} - \frac{x^2}{2}\right) \dd x \dd y = 0,\\
    \int\limits_{-b/2}^{b/2}\int\limits_{-a/2}^{a/2} y\Phi \dd x \dd y = \int\limits_{-b/2}^{b/2}\int\limits_{-a/2}^{a/2} \Phi_{,y}\left(\frac{b^2}{8} - \frac{y^2}{2}\right) \dd x \dd y = 0.
\end{align}
\end{subequations}

\subsection{Equations of motion and their solutions: dispersion relation and displacement field of low-energy acoustic phonons} \label{subsec:EoM}

In this section, we express the effective Lagrangian for the nanowire, derive the equations of motions from the Lagrangian, and solve those equations. 

The Lagrangian is shown in Eq.~\eqref{eq:thin_lagrangian}.
Using the Fourier decomposition in Eq.~\eqref{eq:straindecomp}, we decompose the Lagrangian as a sum for wavenumbers $q$:
\begin{equation}
    \mathcal{L} = \sum_q \mathcal{L}^q.
\end{equation}
Here, the component of the Lagrangian, corresponding to wavenumber $q$ reads as:
\begin{align} 
    \mathcal{L}^q(\mathbf{v}, \phi,& \boldsymbol{\vartheta}, \eta, \Dot{\mathbf{v}}, \Dot{\phi}, \Dot{\boldsymbol{\vartheta}}, \Dot{\eta}) =\nonumber \\
    \frac{L}{2}\big\{-q^2E&\big[Av_{z}^2 + I_y\vartheta_{x}^2 + I_x\vartheta_{y}^2 + D_1 \eta^2\big] \nonumber \\
     +C_{66}&\big[A(iqv_{x} + \vartheta_x)^2 + D_2 \eta^2 - I_x q^2\phi^2 +  2iq D_4\eta\phi\big] \nonumber\\ 
     +C_{44}&\big[A(iqv_{y} + \vartheta_y)^2 + D_3 \eta^2 - I_y q^2\phi^2 + 2iqD_5\eta\phi\big] \nonumber\\
    - \rho&\big[A\dot{\mathbf{v}}^2 + I\dot{\phi}^2 + I_y\dot{\vartheta}_x^2 + I_x\dot{\vartheta}_y^2 + D_1 \dot{\eta}^2
    \big]\big\}. \label{eq:Lq}
\end{align}
To obtain this form, we used Eqs.~\eqref{eqs:warpingsymmetries} and \eqref{eqs:zero_integrals}, and introduced the following notations:
\begin{subequations}\label{eqs:rigidity}
\begin{eqnarray}
    C &=& 4\mu\int\limits_{-\tilde{b}/2}^{\tilde{b}/2}\int\limits_{-\tilde{a}/2}^{\tilde{a}/2} \chi \dd \tilde{x} \dd \tilde{y}, \label{eq:torsional_rigidity} \\
    D_1 &=& \int\limits_{-b/2}^{b/2}\int\limits_{-a/2}^{a/2} \Phi^2\dd x \dd y,\\ 
    D_{2/3} &=& \int\limits_{-b/2}^{b/2}\int\limits_{-a/2}^{a/2}  \Phi^2_{,x/y}\dd x \dd y,\\ 
    D_4 &=& -\int\limits_{-b/2}^{b/2}\int\limits_{-a/2}^{a/2} y\Phi_{,x}\dd x \dd y,\\ 
    D_5 &=& \int\limits_{-b/2}^{b/2}\int\limits_{-a/2}^{a/2} x\Phi_{,y} \dd x \dd y,
\end{eqnarray}
\end{subequations}

Note that due to the definition of $\chi$ in Eq.~\eqref{eq:chi}, and Eq.~\eqref{eq:twist_shear}, the torsional rigidity, $C$ in Eq.~\eqref{eq:torsional_rigidity}, can be expressed as:
\begin{equation}
    C = C_{66}(D_4 + I_x) + C_{44}(D_5 + I_y), \label{eq:rigid_rel1}
\end{equation}
where $I_x$ and $I_y$ are the moments of inertia defined in Eq.~\eqref{eq:momentofinertia}.
Also, due to the boundary conditions in Eq.~\eqref{eqs:torsional_boundary} and the equation of equilibrium in Eq.~\eqref{eq:torsion_eq}, the following relation holds:
\begin{equation}
    C_{66}(D_2 + D_4) + C_{44}(D_3 + D_5) = 0.\label{eq:rigid_rel2}
\end{equation}
In Eq.~\eqref{eq:Lq} and from now on in this section, we omit the $q$ superscript to simplify the notations. 

The Hamilton variational principle, see e.g.~Sec.~E III.~in Ref.~\cite{Gurtin1973}, implies the equations of motion with zero external force. These equations read as
\begin{subequations}\label{eqs:EoMs}
    \begin{align}
    \rho \Ddot{v}_z &= -q^2Ev_{z}, \label{eq:vz}\\
    \rho I\Ddot{\phi} &=  -(C_{66}I_x + C_{44}I_y)(iq\eta+q^2\phi) + iqC\eta,\label{eq:phi}\\
    \rho D_1\Ddot{\eta} &= -q^2ED_1\eta + (C - C_{66}I_x - C_{44}I_y)(\eta -iq\phi),\label{eq:eta}\\
    \rho \Ddot{v}_x &= C_{66}(iq\vartheta_{x} - q^2v_{x}),\label{eq:vx}\\
    \rho I_y\Ddot{\vartheta}_x& = -q^2EI_y\vartheta_{x} - C_{66}A(iqv_{x} +\vartheta_{x}),\label{eq:thetax}\\
    \rho \Ddot{v}_y &= C_{44}(iq\vartheta_{y} - q^2v_{y}),\label{eq:vy}\\
    \rho I_x\Ddot{\vartheta}_y&= -q^2EI_x\vartheta_{y} - C_{44}A(iqv_{y} +\vartheta_{y}).\label{eq:thetay}
\end{align}
\end{subequations}
To obtain Eq.~\eqref{eq:phi} and Eq.~\eqref{eq:eta}, we have used the relations in Eqs.~\eqref{eq:rigid_rel1} and \eqref{eq:rigid_rel2}.

The set of equations \eqref{eqs:EoMs} can be compactly represented as
\begin{equation}
\label{eq:vectorized-eom}
    \rho \Ddot{\boldsymbol{\alpha}} = -\mathbf{D} \boldsymbol{\alpha},
\end{equation}
with the following notation introduced for the 7-component vector $\boldsymbol{\alpha}$ of generalized coordinates and the $7 \times 7$ dynamical matrix $\mathbf{D}$:
\begin{align} \label{eq:alpha}
    \boldsymbol{\alpha} &=(v_z,\sqrt{I} \phi, \sqrt{D_1}\eta, \sqrt{A} v_x, \sqrt{I_y}\vartheta_x, \sqrt{A} v_y, \sqrt{I_x}\vartheta_y  
    )^T,\\
    \mathbf{D} &= \begin{pmatrix}
        q^2E & 0 & 0 & 0 \\
        0 & \mathbf{D}_\textrm{T} & 0 & 0 \\
        0 & 0 & \mathbf{D}_{\textrm{F},x} & 0 \\
        0 & 0 & 0 & \mathbf{D}_{\textrm{F},y}
    \end{pmatrix}, \\
    \mathbf{D}_\textrm{T} &= \begin{pmatrix}
        q^2\frac{(C_{66}I_x + C_{44}I_y)}{I} & iq \frac{(C_{66}I_x + C_{44}I_y) - C}{\sqrt{ID_1}} \\
        -iq\frac{(C_{66}I_x + C_{44}I_y)-C}{\sqrt{ID_1}} & \frac{(C_{66}I_x + C_{44}I_y) - C}{D_1} + q^2E
    \end{pmatrix} \\
    \mathbf{D}_{\textrm{F},x} &= \begin{pmatrix}
        q^2C_{66} & -iq C_{66}\sqrt{\frac{A}{I_y}} \\
        iq C_{66}\sqrt{\frac{A}{I_y}} & C_{66}\frac{A}{I_y} + q^2E
    \end{pmatrix} \\
    \mathbf{D}_{\textrm{F},y} &= \begin{pmatrix}
        q^2C_{44} & -iq C_{44}\sqrt{\frac{A}{I_x}} \\
        iq C_{44}\sqrt{\frac{A}{I_x}} & C_{44}\frac{A}{I_x} + q^2E
    \end{pmatrix}.
\end{align}
In the definition of $\boldsymbol{\alpha}$, we inserted the factors in front of the generalized coordinates carefully, such that the dynamical matrix $\mathbf{D}$ is Hermitian. Therefore, we can use standard perturbation theory to find its eigenvalues and eigenvectors.

The differential equation \eqref{eq:vectorized-eom} describes a coupled harmonic oscillator system, which we solve with the following harmonic ansatz:
\begin{equation}
    \boldsymbol{\alpha}(t) = \boldsymbol{\alpha}_0 e^{i\omega t}.
\end{equation}
This ansatz provides a solution if $\boldsymbol{\alpha}_0$ is the solution of the eigenvalue equation
\begin{equation}
    -\omega^2 \rho \boldsymbol{\alpha}_0 = \mathbf{D} \boldsymbol{\alpha}_0.
\end{equation}

The dynamical matrix $\mathbf{D}$ decouples into four independent blocks, corresponding to the four modes: longitudinal, torsional and two flexural modes; which we treat separately. 

The first, $1 \times 1$ block of $\mathbf{D}$ corresponds to the longitudinal mode, which can be trivially solved. Using the definition of $\boldsymbol{\alpha}$ in Eq.~\eqref{eq:alpha} and substituting the generalized coordinates into Eq.~\eqref{eq:rod_displacement}, we obtain the results in terms of the displacement field and dispersion relation. These read as
\begin{subequations}\label{eq:longidudinal}
    \begin{eqnarray}
        \mathbf{u} &=& A_\textrm{L}[\hat{\mathbf{z}} -iq(\nu_x x\hat{\mathbf{x}} + \nu_y y\hat{\mathbf{y}})] e^{i(qz - \omega_\textrm{L}t)},\\
        \omega_\textrm{L}(q) &=&  \sqrt{\frac{E}{\rho}} q,
    \end{eqnarray}
\end{subequations}
where $A_\textrm{L}$ is a constant specifying the amplitude of the wave, and the first-order correction in $qd$ is calculated from Eqs.~\eqref{eq:warping_zeroavg} and \eqref{eqs:stress_constraint}. Higher-order corrections are neglected. 

The next $2\times 2$ block $\mathbf{D}_\mathrm{T}$ of $\mathbf{D}$ corresponds to the torsional mode. 
We solve the eigenvalue equation of $\mathbf{D}_\textrm{T}$ perturbatively in $q$. We only calculate the mode that has zero eigenvalue in zeroth order. The solution, up to first order in $q$ for $\boldsymbol{\alpha}$ and second order in $q$ for $\omega^2$, reads as
\begin{equation}
    \boldsymbol{\alpha}_0 = \phi_0\sqrt{I}\begin{pmatrix}
        0, 1, iq\sqrt{\frac{D_1}{I}}, 0, 0, 0, 0
    \end{pmatrix}^T, \quad \omega^2 = q^2\frac{C}{I\rho},
\end{equation}
where $\phi_0$ is an arbitrary scalar and $I = I_x + I_y$.
We obtain the corresponding displacement field and dispersion relation using  Eq.~\eqref{eq:rod_displacement}.
These read as
\begin{subequations}\label{eq:torsional}
    \begin{eqnarray}
        \mathbf{u} &=& \phi_{0}\left[\hat{\mathbf{z}} \cross \boldsymbol{\varrho} + iq\Phi(x,y)\hat{\boldsymbol{z}}  \right]e^{i(qz - \omega_\textrm{T}t)},\\
        \omega_\textrm{T}(q) &=&  \sqrt{\frac{C}{\rho I}} q.
    \end{eqnarray}
\end{subequations}

The third and fourth $2\times 2$ blocks $\mathbf{D}_{\mathrm{F},x}$
and
$\mathbf{D}_{\mathrm{F},y}$
correspond to the $x$ directional and $y$ directional flexural modes, respectively. We solve the eigenvalue problem perturbatively, for the modes which have zero eigenvalue in zeroth order.
Note that the first, second, and third order corrections are all zero in this case. 
To collect all terms up to fourth order in $q$, we do fourth-order perturbation theory for the eigenvalue but for the eigenvector we only keep terms up to second order. 

The results for the $x$ directional flexural mode read
\begin{subequations}
\begin{align}
    \boldsymbol{\alpha}_0 &= A_{\textrm{F},x}\sqrt{A}\begin{pmatrix}
        0, 0, 0, 1 - q^2\frac{I_y}{2A}, -iq\sqrt{\frac{I_y}{A}}, 0, 0, 0
    \end{pmatrix}^T, \\
    \omega^2 &= q^4\frac{EI_y}{A\rho}.
\end{align}
\end{subequations}
The displacement field and  dispersion relation, through Eq.~\eqref{eq:alpha} and \eqref{eq:rod_displacement}, then read as
\begin{subequations}\label{eq:flexuralx}
    \begin{align}
        \mathbf{u} &= A_{\textrm{F},x}\Bigg[\Bigg(1 - q^2\frac{I_y}{2A}\Bigg)\hat{\mathbf{x}} - iqx\hat{\mathbf{z}} - q^2\Bigg(xy\nu_y\hat{\mathbf{y}}\nonumber\\
        &\quad  +\frac{\nu_x (x^2 - I_y) - \nu_y (y^2 - I_x)}{2}\hat{\mathbf{x}}\Bigg)   \Bigg]e^{i(qz - \omega_{\textrm{F},x}t)},\\
        \omega_{\textrm{F},x}(q) &=  \sqrt{\frac{EI_y}{\rho A}} q^2.
    \end{align}
\end{subequations}
In the displacement field, similarly to the longitudinal case, we have introduced non-zero $x$ and $y$ components of $\boldsymbol{\Psi}^q$ term, which reads as
\begin{subequations} \label{eqs:PsiFx}
\begin{eqnarray}
    \Psi^q_x &=& - q^2A_{\textrm{F},x}\frac{\nu_x (x^2 - I_x) - \nu_y (y^2 - I_y)}{2},\\
    \Psi^q_y &=& -q^2A_{\textrm{F},x}xy\nu_y.
    \end{eqnarray}
\end{subequations}
This correction is incorporated in order to satisfy assumption 2, and it is determined using Eqs.~\eqref{eqs:stress_constraint} and \eqref{eq:warping_zeroavg}.

Similarly, for the $y$ directional mode, the displacement field reads as
\begin{subequations}\label{eq:flexuraly}
    \begin{align}
        \mathbf{u} &= A_{\textrm{F},y}\Bigg[\Bigg(1 - q^2\frac{I_x}{2A}\Bigg)\hat{\mathbf{y}} - iqy\hat{\mathbf{z}} - q^2\Bigg(xy\nu_x\hat{\mathbf{x}}\nonumber\\
        &\quad  +\frac{\nu_y (y^2 - I_x) - \nu_x (x^2 - I_y)}{2}\hat{\mathbf{y}}\Bigg)   \Bigg]e^{i(qz - \omega_{\textrm{F},y}t)},\\
        \omega_{\textrm{F},y}(q) &=  \sqrt{\frac{EI_x}{\rho A}} q^2.
    \end{align}
\end{subequations}

\subsection{Discussion of the assumptions and boundary conditions} \label{subsec:assumptions_discussion}

In this section, we review the assumptions we made in Sec.~\ref{subsec:assumptions}, and confirm that they are consistent with the solutions obtained in Sec.~\ref{subsec:EoM}. We also confirm that the boundary conditions stated in Eq.~\eqref{eqs:boundary} hold for the solutions of Sec.~\ref{subsec:EoM}.

Assumption 1 is a restriction of the displacement field. The ansatz in Eq.~\eqref{eq:warping_factorize} is generally used in the literature \cite{Gurtin1973, Vlasov1961Thin-walledBeams, Landau1986TheoryElasticity, Davi1996DynamicsRods}; however, we are unaware of a rigorous and physically justified argument for it. 
Nevertheless, the high-precision agreement between the resulting analytical dispersion relation with the numerical results, shown in Fig.~\ref{fig:dispersion}, as well as the analytical-numerical comparison in Fig.~\ref{fig:gammamap}(b), are reassuring, and provide strong empirical justification for applying the assumption. 

Assumption 2 holds exactly for all the solutions we have found. However, in the case of the longitudinal and flexural modes, we had to introduce non-zero $x$ and $y$ components of $\boldsymbol{\Psi}^q$, therefore we need to check that the introduced terms are consistent with Assumption 3. For the torsional mode these components are zero.

In case of the longitudinal mode, assumption 2 holds exactly. However, it results in nonzero shear stress, which is calculated from Eq.~\eqref{eq:longidudinal}:
\begin{equation} \label{eq:shearL}
    \sigma^q_{xz} = q^2A_\textrm{L} C_{66}\nu_x xe^{-i\omega_\textrm{L}t},\quad \sigma^q_{yz} = q^2A_\textrm{L} C_{44} \nu_y y e^{-i\omega_\textrm{L}t}.
\end{equation}
These components are second order in $q$, while the solution is only first order, and therefore these contributions are negligible, which is the statement of assumption 3. 

Similarly, for the flexural modes, the $x$ and $y$ components of $\boldsymbol{\Psi}^q$ are non-zero. 
For the $x$ directional mode, these read as in Eqs.~\eqref{eqs:PsiFx}.
This form of the displacement field ensures that assumption 2 holds exactly.
However, the resulting nonzero shear stress components are of third order in $q$, and therefore negligible, and in this sense, Assumption 3 is consistent with the solution. Similar analysis holds for the $y$ directional flexural mode.

Let us now check whether the boundary conditions in Eq.~\eqref{eqs:boundary} hold for the solutions. The conditions for $\sigma_{xx}$, $\sigma_{xy}$ and $\sigma_{yy}$ are satisfied due to Assumption (2), which holds exactly for all of the solutions. Therefore, only the $\sigma_{xz}$ and $\sigma_{yz}$ components of the stress tensor needs to be checked. 

The shear stress components were already checked for the longitudinal and flexural modes, because the only non-zero contributions in these case are due to the $x$ and $y$ components of $\boldsymbol{\Psi}^q$, and these are negligible.

For the torsional mode, the shear stress reads as
\begin{subequations}
\begin{align}
    \sigma^q_{xz} &= \frac{iqC_{66}\phi_0}{2} (\Phi_{,x} - y)e^{-i\omega_\textrm{T}t},\\
    \sigma^q_{yz} &= \frac{iq\phi_0}{2} (\Phi_{,y} + x)e^{- i\omega_\textrm{T}t},
\end{align}
\end{subequations}
and these are zero on the corresponding boundaries due to the property of the warping function, see Appendix~\ref{subsec:warping}. 

In conclusion, the assumptions we made prior to the derivation are consistent with the solutions we obtained. In addition, the solutions satisfy the boundary conditions.

\section{Quantization of the displacement} \label{app:phononQuantization}
In this section, we briefly summarize and derive the quantization of the displacement following the standard method \cite{Solyom2007FundamentalsSolids}. Let us assume that we have the solution for Eq.~\eqref{eq:reducedChristoffel}. Then, the displacement field can be decomposed as the sum of the eigenmodes as
\begin{equation}
    \mathbf{u}(\mathbf{r}, t) = \sum_{\lambda, q} \mathbf{w}_\lambda^{q} (x,y) e^{i(qz -\omega t)} = \sum_{\lambda, q} Q_\lambda (q,t) \mathbf{e}_\lambda^{q}(x,y)e^{iqz
    },
\end{equation}
where $\mathbf{e}_\lambda^{q}(x,y)$ is normalized as
\begin{equation}
    \int\limits_A |\mathbf{e}_\lambda^{q}(x,y)|^2 \dd x\dd y = A.
\end{equation}
The elastic energy reads as \cite{Landau1986TheoryElasticity}
\begin{equation}
    U = \frac{1}{2}\int\limits_V \sum_{ijkl} \frac{\partial u_i}{\partial r_j}C_{ijkl} \frac{\partial u_k}{\partial r_l} = - \frac{1}{2}\int\limits_V \sum_{ijkl} u_i C_{ijkl} \frac{\partial^2 u_k}{\partial r_j\partial r_l}.
\end{equation}
Using the eigenmode decomposition it reads as
\begin{equation}
    U = \frac{1}{2}\sum_{\lambda,q}  AL \rho \omega^2_\lambda(q) |Q_\lambda(q)|^2.
\end{equation}
The kinetic term reads as
\begin{equation}
     \frac{1}{2}\rho \int\limits_V \left(\frac{\dd \mathbf{u}}{\dd t}\right)^2  =  \frac{1}{2}\sum_{\lambda,q}\rho A L |\dot{Q}_\lambda (q)|^2,
\end{equation}
and the Lagrangian as
\begin{equation}
    \mathcal{L} = \frac{1}{2}\sum_{\lambda,q} \rho A L|\dot{Q}_\lambda (q)|^2-\rho A L\omega^2_\lambda(q) |Q_\lambda(q)|^2.
\end{equation}
The canonical momentum corresponding to $Q$ is then
\begin{equation}
    P_\lambda (q) = \frac{\partial \mathcal{L}}{\partial \dot{Q}_\lambda (q)} = L\rho A \dot{Q}^*_\lambda(q).
\end{equation}
We can now construct the Hamiltonian as
\begin{equation}
    \mathrm{H} = \frac{1}{2}\sum_{\lambda,q} \frac{|P_\lambda(q)|^2}{LA\rho} + L\rho A\omega^2_\lambda(q) |Q_\lambda(q)|^2.
\end{equation}
After canonical quantization, the annihilation operators are introduced as
\begin{equation}
    a_\lambda(q) = \sqrt{\frac{LA\rho \omega_\lambda (q)}{2\hbar}}\left(Q_\lambda(q) + \frac{i}{LA\rho \omega_\lambda (q)} P_\lambda (q) \right),
\end{equation}
and the inverse relation reads as
\begin{subequations}
\begin{align}
    Q_\lambda (q) = \frac{Q_0^{\lambda}(q)}{\sqrt{2}}\left(a_\lambda (q) + a^\dagger_\lambda (-q) \right), \\
    Q_0^{\lambda} (q) =  \sqrt{\frac{\hbar}{LA\rho \omega_\lambda (q)}}
\end{align}
\end{subequations}
and the displacement field expressed with the creation and annihilation operators results in Eq.~\eqref{eq:displacement}.

\bibliography{manual, ssq}

\end{document}